\RequirePackage{rotating}
\documentclass[fleqn,usenatbib]{mnras}

\usepackage[T1]{fontenc}

\DeclareRobustCommand{\VAN}[3]{#2}
\let\VANthebibliography\thebibliography
\def\thebibliography{\DeclareRobustCommand{\VAN}[3]{##3}\VANthebibliography}

\usepackage{graphicx}	
\usepackage{amsmath}	
\usepackage{longtable}
\usepackage[graphicx]{realboxes}
\usepackage{rotating}
\usepackage{lscape}
\usepackage{parskip}
\PassOptionsToPackage{hidelinks}{hyperref}

\usepackage{newtxtext,newtxmath}

\title[Three new hot Jupiters, TOI-1181b, TOI-1516b and TOI-2046b]{TOI-2046b, TOI-1181b and TOI-1516b, three new hot Jupiters from \textit{TESS}:
planets orbiting a young star, a subgiant and a normal star}

\author[Kab\'{a}th, P. et al.]{Petr Kab\'{a}th,$^{1}$\thanks{E-mail: petr.kabath@asu.cas.cz}
Priyanka Chaturvedi,$^{2}$
Phillip J. MacQueen,$^{3}$
Marek Skarka,$^{1}$
J\'{a}n \v{S}ubjak ,$^{1,4}$\newauthor
Massimilliano Esposito,$^{2}$
William D. Cochran,$^{9}$
Salvatore E. Bellomo,$^{11}$
Raine Karjalainen,$^{1}$
Eike W. Guenther,$^{2}$\newauthor
Michael Endl,$^{10}$      
Szil\'ard Csizmadia,$^{5}$
Marie Karjalainen,$^{1}$
Artie Hatzes,$^{2}$
Ji\v{r}\'{i} \v{Z}\'{a}k,$^{6}$
Davide Gandolfi,$^{11}$ \newauthor
Henri M.J. Boffin,$^{6}$
Jose I. Vines,$^{7}$
John H. Livingston,$^{8}$
Rafael A. Garc\'{i}a$^{35}$
Savita Mathur$^{12,13}$ \newauthor
Luc\'{i}a Gonz\'{a}lez-Cuesta$^{12,13}$
Martin Bla\v{z}ek,$^{1,21}$ 
Douglas A. Caldwell,$^{23}$
Knicole D. Col\'{o}n,$^{24}$
Hans Deeg,$^{12,13}$ \newauthor    
Anders Erikson,$^{5}$
Vincent Van Eylen,$^{15}$
William Fong,$^{26}$
Malcolm Fridlund,$^{16,17}$
Akihiko Fukui,$^{29,12}$\newauthor
G{\' a}bor~F{\H u}r{\' e}sz,$^{26}$
Robert F. Goeke,$^{25}$
Elisa Goffo,$^{11,2}$
Steve Howell,$^{22}$
Jon M. Jenkins,$^{22}$
Peter Klagyivik,$^{5}$\newauthor
Judith Korth,$^{18}$
David W. Latham,$^{18}$
Rafael Luque,$^{27}$
Dan Moldovan,$^{28}$
Felipe Murgas,$^{12}$
Norio Narita,$^{29,30,31,12}$\newauthor
Jaume Orell-Miquel,$^{12,13}$
Enric Palle,$^{12,13}$
Hannu Parviainen,$^{12,13}$
Carina M. Persson,$^{16}$
Phillip A. Reed,$^{14}$\newauthor
Seth Redfield,$^{19}$
George R. Ricker,$^{25}$ 
Sara Seager,$^{26,33,34}$
Luisa Maria Serrano,$^{11}$
Avi Shporer,$^{26}$\newauthor
Alexis M. S. Smith,$^{5}$
Noriharu Watanabe,$^{32}$
Joshua N.\ Winn,$^{20}$
and the KESPRINT team
\\
\\
$^{1}$Astronomical Institute of the Czech Academy of Sciences, Fri\v{c}ova 298, 25165, Ond\v{r}ejov, Czech Republic\\
$^{2}$Thueringer Landessternwarte Tautenburg, Sternwarte 5, 07778 Tautenburg, Germany\\
$^{3}$McDonald Observatory, The University of Texas at Austin, Austin, TX 78712, USA\\
$^{4}$Astronomical Institute of Charles University, V Hole\v{s}ovi\v{c}k\'ach 2, 180 00, Prague, Czech Republic\\
$^{5}$Department of Extrasolar Planets and Atmospheres, German Aerospace Center, Rutherfordstrasse 2., D-12489 Berlin, Germany\\
$^{6}$European Southern Observatory, Karl-Schwarzschild-Str. 2., D-85748 Garching bei Muenchen, Germany\\
$^{7}$Departamento de Astronom\'ia,
Universidad de Chile, Casilla 36-D, Santiago, Chile\\
$^{8}$Department of Astronomy, University of Tokyo, 7-3-1 Hongo, Bunkyo-ku, Tokyo 113-0033, Japan\\
$^{9}$Center for Planetary Systems Habitability and McDonald Observatory, The University of Texas, Austin, TX 78712, USA\\
$^{10}$Center for Planetary Systems Habitability and Department of Astronomy, The University of Texas, Austin, TX 78712, USA\\
$^{11}$Dipartimento di Fisica, Università Degli Studi di Torino
Via Pietro Giuria, 1, 10125 Torino\\
$^{12}$Instituto de Astrof\'isica de Canarias (IAC), 38205 La Laguna, Tenerife, Spain\\
$^{13}$Departamento de Astrof\'isica, Universidad de La Laguna (ULL), 38206 La 
Laguna, Tenerife, Spain\\
$^{14}$Department of Physical Sciences, Kutztown University, Kutztown, PA, 19530, USA\\
$^{15}$Mullard Space Science Laboratory, University College London, Holmbury St Mary, Dorking, Surrey RH5 6NT, UK\\
$^{16}$Chalmers University of Technology, Department of Space, Earth and Environment, Onsala Space Observatory,  SE-439 92 Onsala, Sweden\\
$^{17}$Leiden Observatory, University of Leiden, PO Box 9513, 2300 RA, Leiden, The Netherlands \\
$^{18}$ Department of Space, Earth and Environment, Astronomy and Plasma Physics, Chalmers University of Technology, 412 96 Gothenburg, Sweden\\
$^{19}$Astronomy Department and Van Vleck Observatory, Wesleyan
University, Middletown, CT 06459, USA\\
$^{20}$ Department of Astrophysical Sciences, Peyton Hall, 4 Ivy Lane,
Princeton, NJ 08544, USA\\
$^{21}$ Department of Theoretical Physics and Astrophysics, Faculty of Science, Masaryk University, Kotl\'{a}\v{r}sk\'{a} 267/2, 611 37~~Brno, Czech 
Republic\\
$^{22}$ NASA Ames Research Center, Moffett Field, CA 94035, USA\\
$^{23}$ SETI Institute, 189 Bernardo Ave, Suite 200 Mountain View, CA 94043, USA\\
$^{24}$ NASA Goddard Space Flight Center, Greenbelt, MD 20771, USA\\
$^{25}$ Kavli Institute for Astrophysics and Space Research, Massachusetts Institute of Technology, Cambridge, MA 02139, USA\\
$^{26}$ Department of Physics and Kavli Institute for Astrophysics and Space
Research, Massachusetts Institute of Technology, Cambridge, MA 02139, USA\\
$^{27}$ Instituto de Astrof\'isica de Andaluc\'ia (IAA-CSIC), Glorieta de la
Astronom\'ia s/n, 18008 Granada, Spain\\
$^{28}$ Google, Cambridge, MA, USA\\
$^{29}$ Komaba Institute for Science, The University of Tokyo, 3-8-1 Komaba,
Meguro, Tokyo 153-8902, Japan\\
$^{30}$ Japan Science and Technology Agency, PRESTO, 3-8-1 Komaba, Meguro, Tokyo
153-8902, Japan\\
$^{31}$ Astrobiology Center, 2-21-1 Osawa, Mitaka, Tokyo 181-8588, Japan\\
$^{32}$ Department of Multi-Disciplinary Sciences, Graduate School of Arts and
Sciences, The University of Tokyo, 3-8-1 Komaba, Meguro, Tokyo 153-8902,
Japan\\
$^{33}$Department of Earth, Atmospheric and Planetary Sciences, Massachusetts
Institute of Technology, Cambridge, MA 02139, USA\\
$^{34}$Department of Aeronautics and Astronautics, MIT, 77 Massachusetts
Avenue, Cambridge, MA 02139, USA\\
$^{35}$Laboratoire AIM, CEA/DSM – CNRS – Université Paris Diderot – IRFU/SAp, 91191
Gif-sur-Yvette Cedex, France
}

\date{Accepted XXX. Received YYY; in original form ZZZ}

\pubyear{2021}

\begin{document}
\pagerange{\pageref{firstpage}--\pageref{lastpage}}
\label{firstpage}
\maketitle
\clearpage

\begin{abstract}
We present the confirmation and characterization of three hot Jupiters, TOI-1181b, TOI-1516b, and TOI-2046b, discovered by the TESS space mission. The reported hot Jupiters have orbital periods between 1.4 and 2.05 days. The masses of the three planets are $1.18\pm0.14$ M$_{\mathrm{J}}$, $3.16\pm0.12$\, M$_{\mathrm{J}}$, and 2.30 $\pm 0.28$  M$_{\mathrm{J}}$, for TOI-1181b, TOI-1516b, and  TOI-2046b, respectively.  The stellar host of TOI-1181b is a F9IV star, whereas TOI-1516b and TOI-2046b orbit F main sequence host stars. The ages of the first two systems are in the range of 2-5 Gyrs. However, TOI-2046 is among the few youngest known planetary systems hosting a hot Jupiter, with an age estimate of 100-400 Myrs. The main instruments used for the radial velocity follow-up of these three planets are located at Ond\v{r}ejov, Tautenburg and McDonald Observatory, and all three are mounted on 2-3 meter aperture telescopes, demonstrating that mid-aperture telescope networks can play a substantial role in the follow-up of gas giants discovered by \textit{TESS} and in the future by \textit{PLATO}. \newline
\end{abstract}

\begin{keywords}
techniques: spectroscopic -- techniques: radial velocities -- planets and satellites: detection -- stars: individual: TOI-1181, TOI-1516, TOI-2046
\end{keywords}



\section{Introduction}

Hot Jupiters are planets with a mass and radius similar to Jupiter while orbiting their host star in less than about $10$ days. Due to their short proximity to their stellar host, hot Jupiters belong to an extremely puzzling group of planets from the evolutionary point of view. Already the first hot Jupiter ever discovered, 51 Peg\,b, raised the question of the formation and evolution of such systems \citep{1995Natur.378..355M}. Immediately, some classical theories of gas planet formation were challenged \citep{1996Icar..124...62P}. 

In spite of over 25 years of studies of  hot Jupiters, there are at least two very good reasons 
to continue such work:\newline

1) insights into planet formation\newline
2) atmospheric characterization\newline

Although hot Jupiters were the first  exoplanets to be discovered, this is an observational bias due to their short orbital periods and large sizes. Exoplanet surveys later showed that the occurrence rate of hot Jupiters is about $0.43\%$ \citep{2017AJ....153..187M} -- less than the $5.2\%$ fraction of Jupiter-sized planets on longer orbits and significantly lower than the occurrence rates of smaller planets \citep{2013ApJ...766...81F}. However, the occurrence rate of giant planets differs also based on the metallicity of the host star, with metal-rich stars being more likely to host a giant planet \citep{Wang_2015}. 

It is still not clear whether hot Jupiters formed in-situ \citep{2016ApJ...829..114B} or if they formed far away from their host star and migrated inwards \citep{2019A&A...628A..42H}. Furthermore, systems with several Jupiter-sized planets such as HR~8799 are most likely very rare \citep{2009ApJ...693.1084W, 2008Sci...322.1348M}. This particular system, HR~8799, is also an example of a young exoplanetary system of an age of a few hundreds of millions of years \citep{2001ApJ...546..352S}. Only a handful of such young systems are known. The youngest fully characterized gas giant in terms of mass and radius is the recently discovered \textit{TESS} planet, HIP 67522 b, of about one Jupiter mass with an age of $17\pm 2$ Myrs \citep{2020AJ....160...33R}. Young hot Jupiters are important objects for understanding the evolution of exoplanetary systems as well as the formation history of systems with smaller planets \citep{2017AJ....153..265W,2020ApJ...892L...7H}.  

Hot Jupiters are also interesting targets for investigating exo-atmospheres, as the atmospheric signatures of large gas planets can be detected even with mid-sized aperture telescopes from the ground \citep{2015A&A...577A..62W,2019AJ....158..120Z}. In some cases, the atmospheres of hot Jupiters might be extended and extremely hot \citep{2021A&A...645A..22Y,Cauley_2021}. Nevertheless, atmospheric processes in hot Jupiters are now being understood at a rapid pace and new models are calculated to understand the composition of atmospheres of giant planets \citep{2020NatAs...4..951G}. 

Finally, one should also mention here the aspect of telescope time optimization during the ground-based follow-up of exoplanetary candidates. Hot Jupiters are indeed ideal targets for mid-aperture telescopes equipped with precise spectrographs. Such instruments are capable of performing the necessary confirmation and in some cases, even allow their characterization. Mid-aperture class telescopes are ideal instruments to exclude false positives for transit detections. The typical global rate of false positives in the \textit{Kepler} sample is about $6-7\%$, up to $17\%$ for giant planets only \citep{2013ApJ...766...81F}. The \textit{TESS} mission global false positives rate is about $25\%$ \citep{2021ApJS..254...39G}. Therefore, the confirmation of transit candidates requires many hours of telescope time.

In this paper, we report the confirmation of the \textit{TESS} discovery of three gas giant planets on short orbits. Planetary candidates TOI-1516b, TOI-1181b and TOI-2046b were confirmed via a coordinated radial-velocity (RV) follow-up monitoring at the Ond\v{r}ejov 2m Perek telescope \citep{2020PASP..132c5002K}, at the Alfred Jensch 2m telescope in Tautenburg~\citep{2003EAEJA.....6093H}, and at the 2.7m telescope at McDonald Observatory~\citep{Tull_1995}, as part of a joint KESPRINT effort. 

The KESPRINT\footnote{http://kesprint.science} consortium has been a major contributor for the characterization of gas planets \citep{2019AcA....69..135S,2018MNRAS.481..596J,2018MNRAS.475.1765B,2019MNRAS.484.3522H,2017AJ....153..130E,2017MNRAS.464.2708S,2016AJ....152..132G,2016AJ....151..171J} found by the \textit{K2} mission \citep{2014PASP..126..398H}, and currently also for small planets \citep{2019ApJ...876L..24G, 2019A&A...623A.165E,2021A&A...645A..41L,2020MNRAS.498.4503F,2020AJ....160..114C,2020A&A...639A.132B,2021MNRAS.507.2154V} found by the \textit{TESS} mission \citep{2015JATIS...1a4003R}. Besides small planets, KESPRINT also led the characterization of the first transiting brown dwarf found by \textit{TESS}, TOI-503b \citep{2020AJ....159..151S}.

\section{Observational data sets}\label{sec2}

Here, and in the following section, we describe the  data flow process from the \textit{TESS} discovery light curves up to the RV follow-up observations.

\subsection{\textit{TESS} light curves}\label{1516}

\textit{TESS} is a mission designed to detect exoplanets from space, using the transit method \citep{ricker14}. The spacecraft consists of four identical cameras with four 2k$ \times $2k CCDs monitoring a field-of-view of $24^{\circ}\times24^{\circ}$. The mission covers about $85\%$ of the sky. Until July 2020, TESS recorded every 2 minutes the flux of over 200,000 main-sequence stars. TESS also obtained full-frame images (FFIs) of the entire, four camera field-of-view at a cadence of 30 minutes. The hot Jupiters reported in this article were observed during the northern hemisphere monitoring campaign in 2019 and 2020. All three systems were identified by the \textit{TESS} Quick Look Pipeline \citep[QLP;][]{2020RNAAS...4..204H,2020RNAAS...4..206H}.

\textit{TESS} observed TOI-1181 in sectors 14, 15, 17, 18, 19, 20, 21, 22, 23, 24, 25 and 26 and 40 and an alert was issued by the \textit{TESS} Science Office on 27 August 2019 \citep{2021ApJS..254...39G}. The resulting Data Validation reports \citep{Twicken:DVdiagnostics2018,Li:DVmodelFit2019} are exceptionally clean and the difference image centroiding locates the source of the transit signatures to within $2.5$ arcsec. There are no nearby stars in the \textit{TESS} Input Catalog (TIC)  sufficiently bright to explain the transit signature as a background eclipsing binary. 

\begin{figure*}

	\includegraphics[height=4.7cm,width=\columnwidth ]{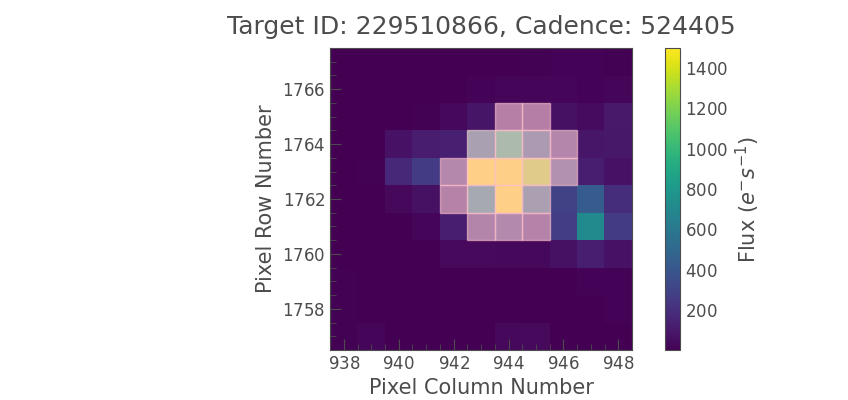}
	\hfill
	\includegraphics[width=\columnwidth]{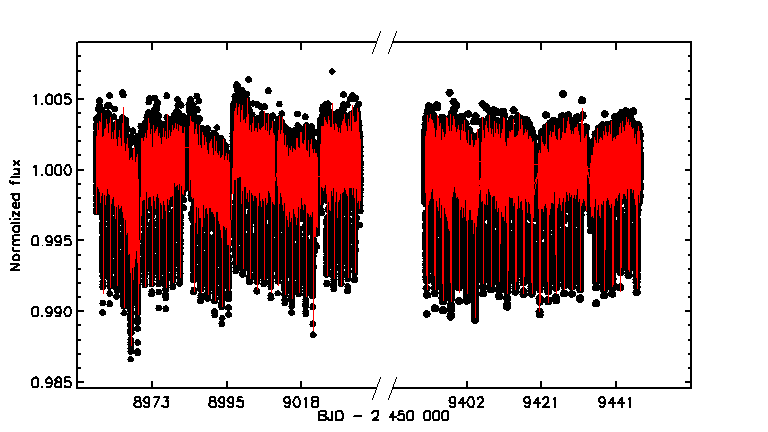}
		\includegraphics[height=4.7cm,width=\columnwidth]{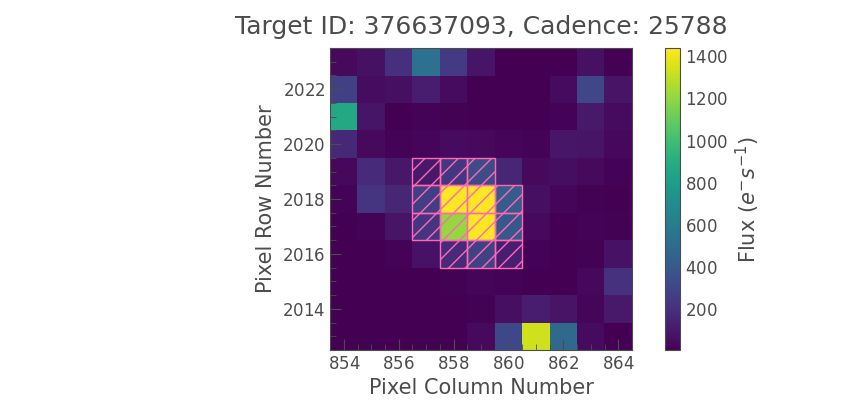}
		\hfill
        \includegraphics[width=\columnwidth]{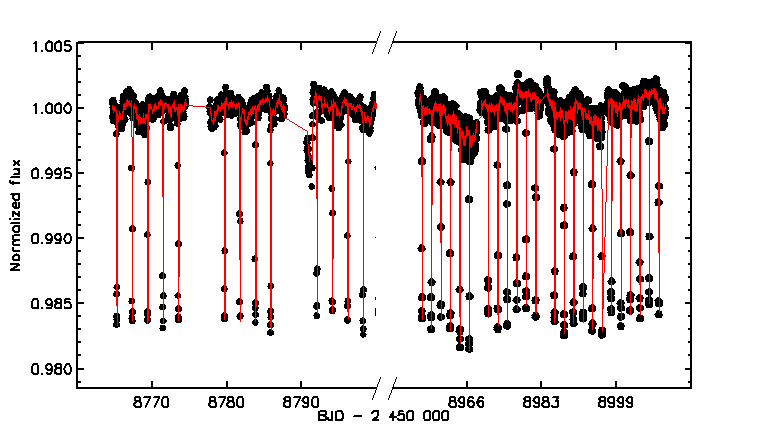}
		\includegraphics[height=4.7cm,width=\columnwidth]{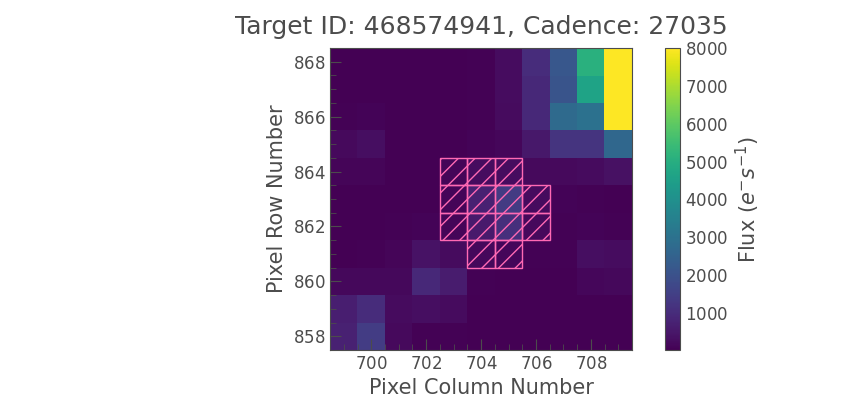}
		\hfill
       \includegraphics[width=\columnwidth]{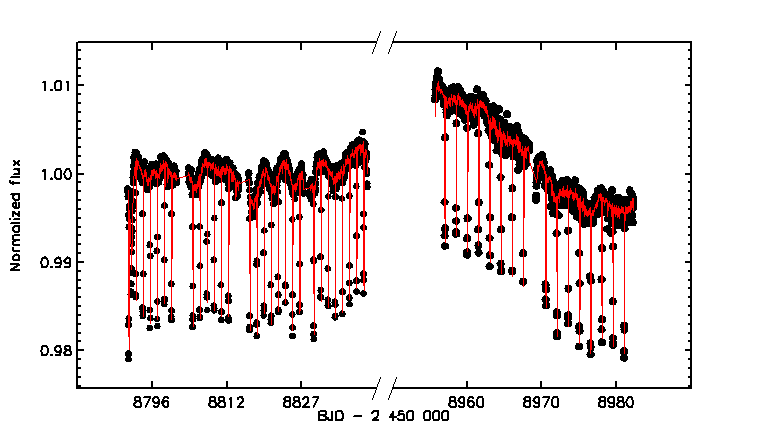}
    \caption{In the left hand panels, the cut-off images from the TPF files are presented with the photometric apertures in pink for stars TOI-1181 (upper panel), TOI-1516 (central panel) and TOI-2046 (bottom panel). The corresponding TESS-SPOC data pipeline processed light curves (SAP, black dots) for the three systems are shown in the right hand panels with the red fitted model obtained with the TLCM code described in Section~\ref{sec3}.} 
    \label{fig1}
\end{figure*}

TOI-1516 was observed in sectors 17, 18, 24 and 25 and an alert was issued on 5 December 2019, while TOI-2046 was observed in sectors 18, 19, 24 and 25 and an alert issued on 19 June 2019. The cutoff images with each target were extracted from the \textit{TESS} Target-Pixel-Files (TPF) with the default optimum aperture using the \textsc{lightkurve} v 2.0 package \citep{2018ascl.soft12013L}. The images are shown in the left-hand panels of Figure \ref{fig1}. In order to inspect initially the \textit{TESS} light curves of the three systems, we used again \textsc{lightkurve 2.0} with the data downloaded from the MAST archive\footnote{https://mast.stsci.edu/portal/Mashup/Clients/Mast/Portal.html}. 

We show the light curves of the three systems obtained with the Simple Aperture Photometry (SAP) pipeline by the Science Processing Operations Center (SPOC) \citep{jenkinsSPOC2016,twicken:PA2010SPIE,2020ksci.rept....6M} in the right-hand side of Figure \ref{fig1}. The full analysis with our noise model description is presented in \ref{subsec:tlcm_fits}.

Catalogue parameters describing the three systems including \textit{TESS}, optical and IR magnitudes and coordinates are summarized in Table~\ref{tab1}.

\begin{table*}
\large
	\centering
	\caption{Observational characteristics of the host stars TOI-1181, TOI-1516 and TOI-2046.}
	\label{tab1}
	\begin{tabular}{lccr} 
		\hline\hline
		System       & TOI-1181 & TOI-1516 & TOI-2046\\
		\hline
		TIC number & TIC 229510866 & TIC 376637093 & TIC 468574941\\
		R.A. & $19^h$ $48^m$ $51^{s}.810$ & $22^h$ $40^m$ $20^{s}.261$ & $01^h$ $04^m$ $44^{s}.362$ \\
		Dec & $+64^{\circ}$ $21^{'}$ $15^{''}.66$ & $+69^{\circ}$ $30^{'}$ $13^{''}.45$ & $+74^{\circ}$ $19^{'}$ $52^{''}.85$\\
		Gaia parallax (mas) & $3.22\pm0.01$ & $4.0540 \pm 0.0098$ & $3.47\pm0.01$ \\
		\textit{TESS}$_\mathrm{mag}$ & $10.079\pm0.009$ & $10.377\pm0.006$ & $10.996\pm0.006$\\
		V$_\mathrm{mag}$ & $10.582 \pm 0.006$& $10.858\pm0.008$ & $11.552\pm0.018$ \\
		Gaia$_\mathrm{mag}$ & $10.4776\pm0.0003$ & $10.7156\pm0.0002$ & $11.4128\pm0.0005$\\
		J$_\mathrm{mag}$ & $9.55\pm0.02$ & $9.92\pm0.03$ & $10.42\pm0.02$\\
		H$_\mathrm{mag}$ & $9.33\pm0.02$ & $9.70\pm0.03$ & $10.13\pm0.03$\\
		K$_\mathrm{mag}$ & $9.22\pm0.02$ & $9.67\pm0.02$ & $10.09\pm0.02$\\
		\textit{WISE}$_{3.4\micron}$ & $9.21\pm0.02$ & $9.60\pm0.02$ & $10.08 \pm0.02$\\
		\textit{WISE}$_{4.6\micron}$ & $9.24\pm0.02$ & $9.63\pm0.02$ & $10.10 \pm0.02$\\
		\textit{WISE}$_{12\micron}$ & $9.24\pm0.02$ & $9.61\pm0.03$ & $10.00 \pm0.05$\\
		\textit{WISE}$_{22\micron}$ & $8.70\pm0.19$ & $9.41\pm0.35$ & $8.88\pm N/A $  \\

		\hline\hline
	\end{tabular}
\end{table*}

\subsection{Ground-based photometric follow-up}\label{1516}

In this section, we describe the ground-based photometric follow-up to check for potential contamination and confirm transit on target. All the presented data were used for a joint fit with the TESS data and with the spectroscopic data sets.

\subsubsection{The Carlson R. Chambliss Astronomical Observatory (CRCAO)}

The Carlson R. Chambliss Astronomical Observatory (CRCAO) is located on the campus of Kutztown University in Pennsylvania and operates a fork-mounted 0.6 m Ritchey–Chrétien optical telescope with a focal ratio of f/8.  The imaging CCD (KAF-6303E) camera has an array of 3k $\times$ 2k (9 $\mu$m) pixels and the system covers a field of view of 19.5' $\times$ 13.0'.  The CCD is cooled to $-30^{\circ}$C.  With $2 \times 2$ binning, the pixel scale is 0.76" and the telescope is auto-guided with a total drift of less than 10 pixels throughout an entire night of continuous observations.  Seeing at CRCAO is typically about 3".

CRCAO observed a full transit of TOI-1181 b on UT 2019 September 25, with 311 consecutive 60 s exposures in the R$_C$ band. However, the observing sequence was interrupted due to a technical problem between fractions of BJD 0.7-0.8, as presented at the top of Figure \ref{lcscrcao}.
CRCAO also observed a full transit of TOI-1516 b on 2020 August 10, with 205 consecutive 90 s exposures in the R$_C$ band.  Both data sets were detrended with airmass. The light curve is shown at the bottom of Figure~ref{lcscrcao}.

\subsubsection{\textit{MUSCAT2} Observations at the Carlos Sanchez Telecope}

The \textit{MUSCAT2} instrument is a four channel imager designed to observe simultaneously in the $g$, $r$, $i$ and $z_s$ photometric bands. It is installed on the 1.52 meter Carlos Sanchez Telescope (TCS) at the Teide Observatory in the Canary Islands, Spain. Each photometric channel is equipped with one 1024x1024 CCD camera with a field of view of 7.4x7.4 arcmin  \citep{Narita2019}. One transit of TOI-1181b was observed with MUSCAT2 on the night of 2019 November 9, using exposure times of 5, 7, 11 and 7 seconds for the $g$, $r$, $i$ and $k_s$ bands, simultaneously. The observations were taken slightly defocused and covered the full transit, with a average seeing trough the night of 1". The resulting light curve is plotted in Fig. \ref{lcmuscat}. Table~\ref{muscat2tab} presents the estimated posterior values of the parameters obtained from the full \textit{MUSCAT2} light curve.

\subsection{High-resolution imaging}\label{1516}

Due to the large pixel scale of the \textit{TESS} detectors, it is important to search for nearby contaminating sources that can dilute the transit depth or lead to false positives \citep[e.g.,][]{Ciardi2015}. In particular, those sources not detected in seeing-limited photometry or by Gaia require specialized observational techniques like high spatial resolution imaging to be detected.

\begin{figure*}
	\includegraphics[width=5.5cm]{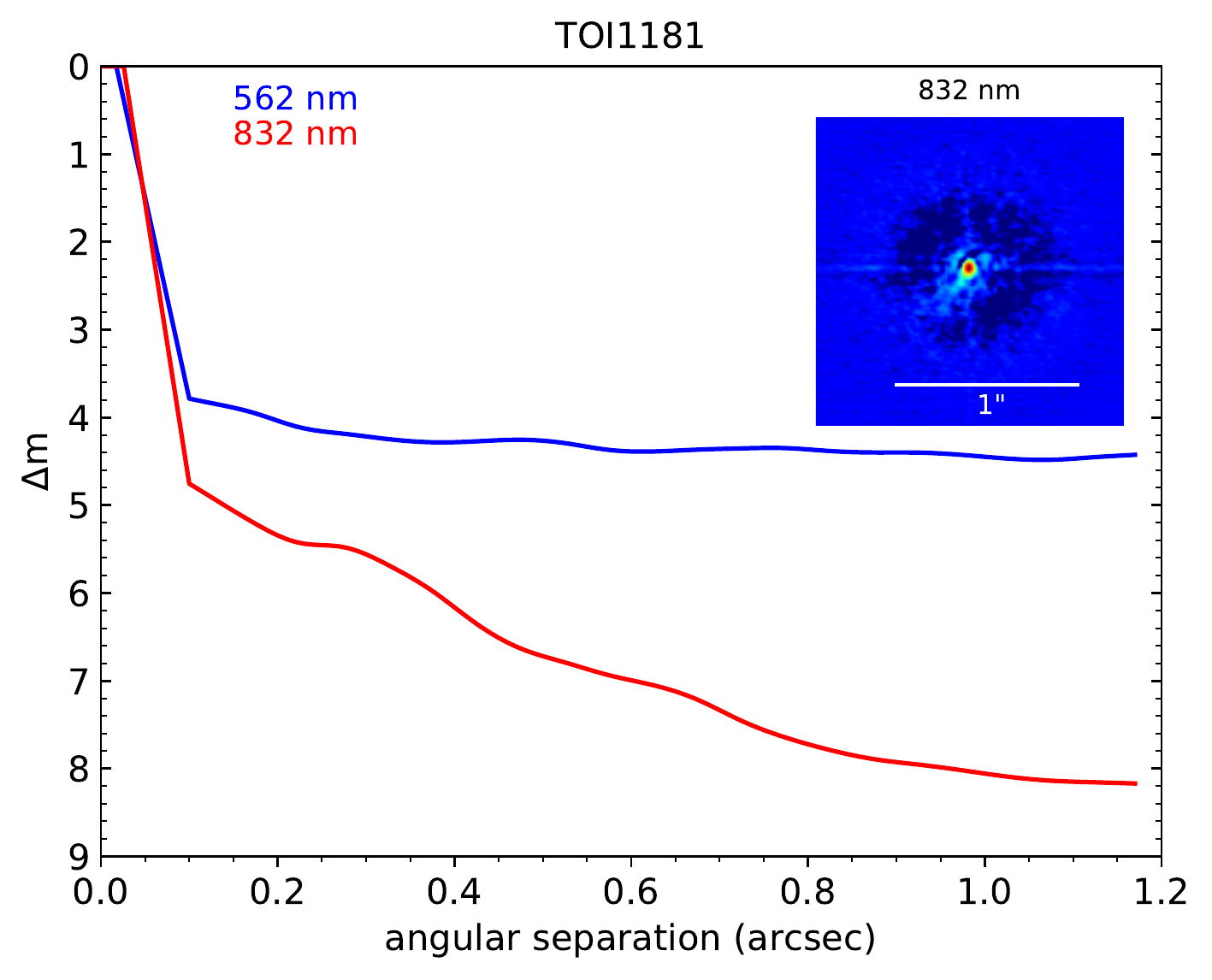}
	\includegraphics[width=5.5cm]{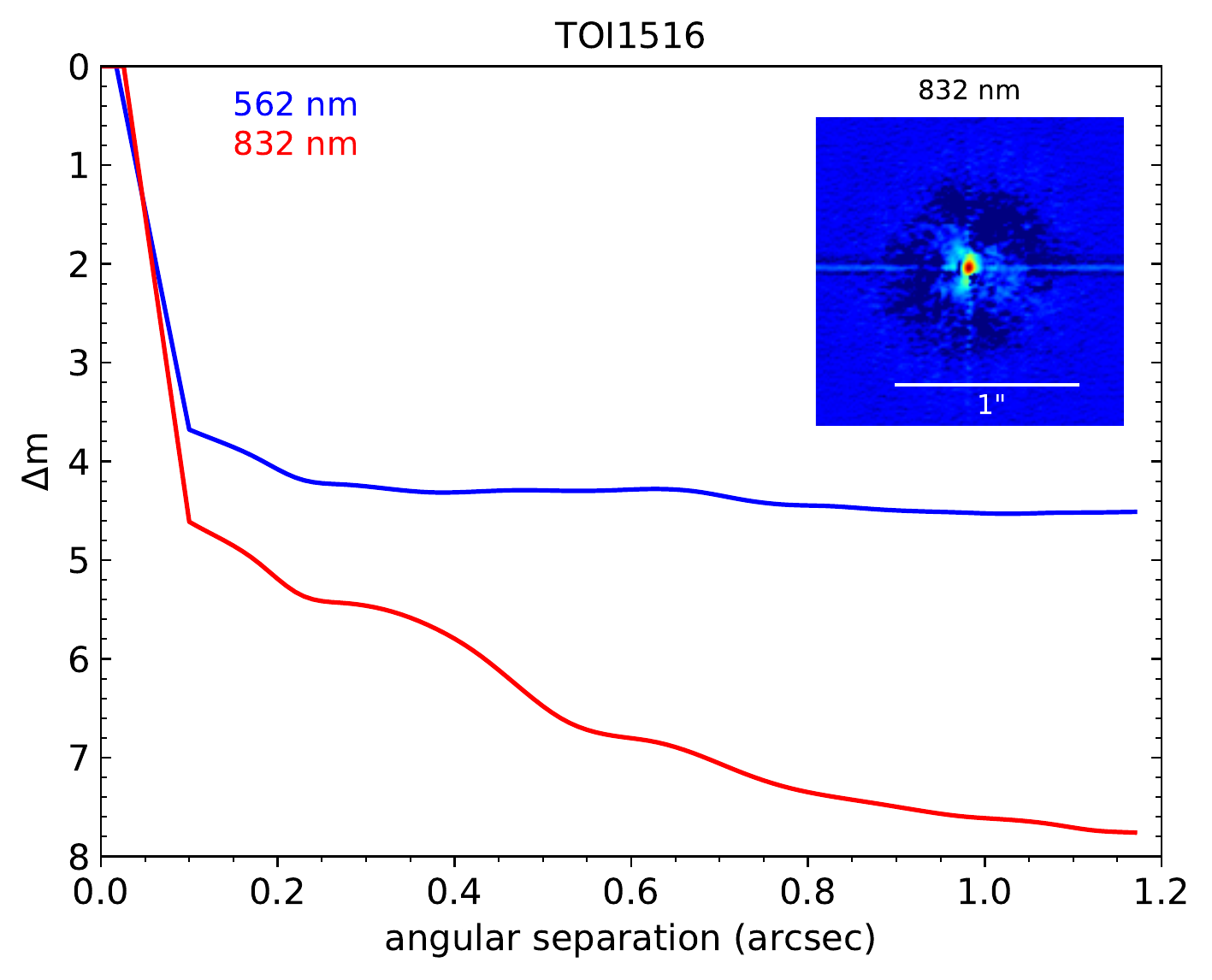}
	\includegraphics[width=5.5cm]{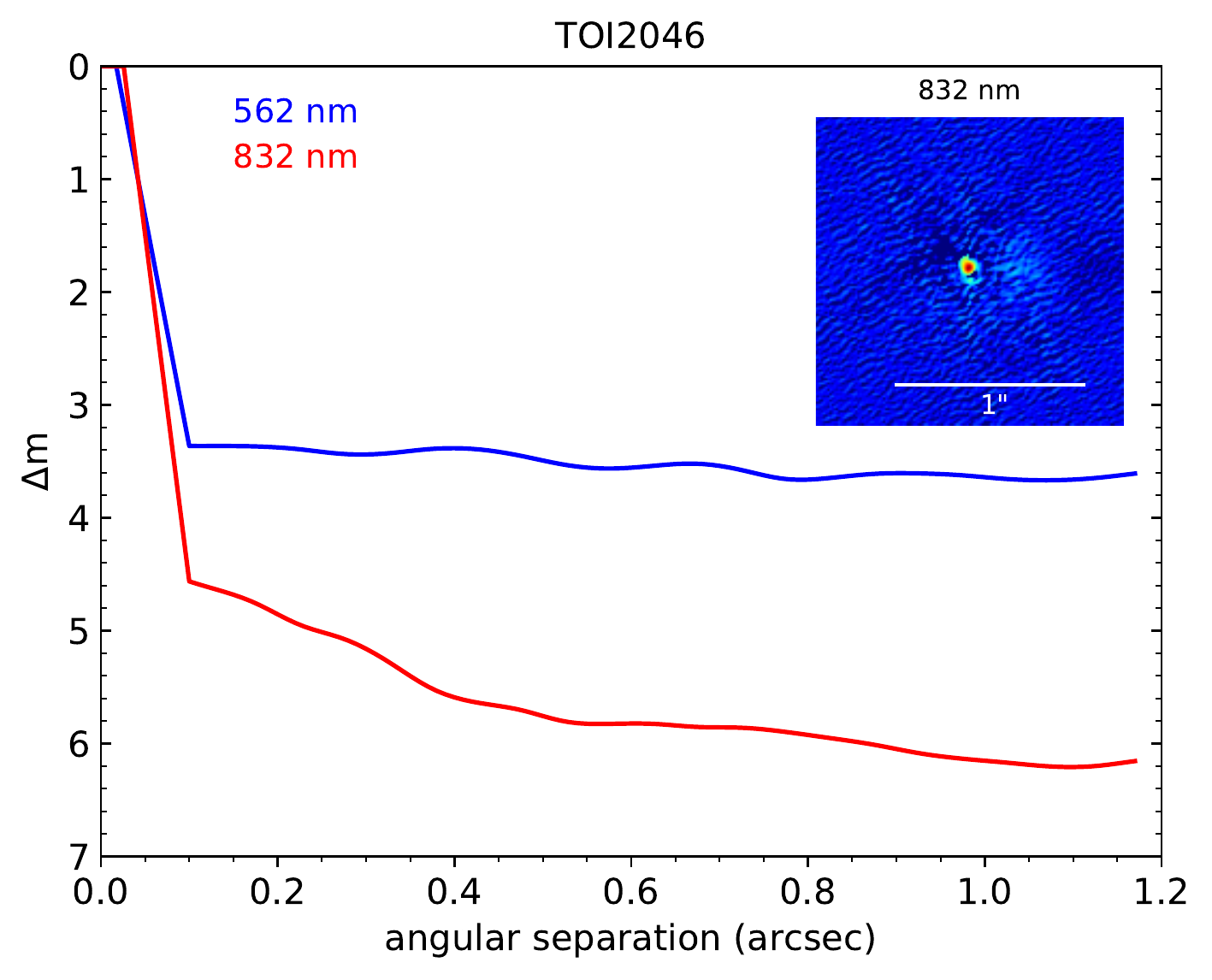}
    \caption{Speckle imaging results of TOI-1181, TOI-1516 and TOI-2046.}
    \label{fig:speckle}
\end{figure*}

On the nights of UT 2020 August 4 and 2021 June 29, we observed TOI-1181, TOI-1516, and TOI-2046 with the 'Alopeke speckle imager \citep{Scott2019}, mounted on the 8.1\,m Gemini North telescope at Mauna Kea. 'Alopeke has a field-of-view of 6.7" and it simultaneously acquires data in two bands centered at 562\,nm and 832\,nm using high-speed electron-multiplying CCDs (EMCCDs). We collected and reduced the data following the procedures described in \citet{Howell2011}. The resulting reconstructed images achieved a contrast of $\Delta\mathrm{mag}\sim4$ at a separation of 0.2\arcsec in the 832\,nm band (see Figure~\ref{fig:speckle}). No contaminants were found.

\subsection{Spectroscopic follow-up}\label{followup}

We performed spectroscopic follow-up using a network of mid-aperture sized telescopes in the northern hemisphere. These include the Ond\v{r}ejov Echelle Spectrograph (OES) located in the Czech Republic, the Tautenburg echelle spectrograph (TCES) in Germany and the Tull Spectrograph at McDonald Observatory in the U.S.A. These provided the bulk of the data for the orbital solutions. All data sets were obtained between January 2020 and March 2021. The measured radial velocities are presented in Tabs. \ref{tabrvs1181}, \ref{tabrvs} and \ref{tab2046}.

\subsubsection{The OES spectrograph}

The Ond\v{r}ejov Echelle Spectrograph (OES) is installed on the 2m telescope in Ond\v{r}ejov, Czech Republic and is operated by the Astronomical Institute of the Czech Academy of Sciences. OES is a fiber-fed spectrograph with resolving power of R=50,000 (at $500$ nm) and wavelength range of 380--900 nm. An accuracy of 70 m/s in radial velocities is typically achieved for a V=6 star. We use a hollow-cathode comparison frame for the wavelength calibration. A standard spectroscopic data reduction with \textsc{IRAF} scripts was performed on all data sets. The instrument related shifts are corrected by telluric lines using the method of  \citet{2006PASP..118..399G} and radial velocities are obtained by the cross-correlation method using the {\tt fxcor} task in \textsc{IRAF}. A more detailed description is provided in \citet{2020PASP..132c5002K}. For the present targets and with an exposure time of 3600s, the typical Signal-to-Noise Ratio (SNR) of the OES spectra range between 20 and 40. 

\subsubsection{The TCES spectrograph}

The Tautenburg Coud\'{e} Echelle spectrograph (TCES) is installed on the 2m Alfred Jensch telescope operated by the Thueringer Landessternwarte Tautenburg, Germany. The instrument achieves a resolving power of R=67,000 over the wavelength range from 467 to 740 nm. The spectra can be calibrated with an iodine (I$_2$) cell. The data reduction is performed with the Tautenburg Spectroscopy Pipeline – $\tau$-spline. The pipeline makes use of standard \textsc{IRAF} and \textsc{PyRaf} routines and the Cosmic Ray code by Malte Tewes based on the method by \citet{2001PASP..113.1420V}. All steps are detailed in, e.g., \citet{2019MNRAS.489.2069S}. The TCES can achieve an RV precision of 3-5 m/s on very bright stars with an iodine cell, but typical RV accuracy on fainter targets is about 70 m/s; for more details, see \citet{2009A&A...507.1659G} and \citet{2009A&A...499..935D}. RVs were derived by a Python-based software Velocity and Instrument Profile EstimatoR (VIPER)\footnote{https://github.com/mzechmeister/viper} \citep{2021ascl.soft08006Z}, based on the standard procedure described in \cite{1996PASP..108..500B,2000A&A...362..585E}. Most of the spectra in Tautenburg were obtained with an iodine cell.
Typical SNRs of the TCES spectra are in the range 10-20 with 1800 sec exposure time.

\subsubsection{The Tull spectrograph}

The Tull spectrograph is installed at the 2.7 m telescope of the McDonald Observatory operated by the University of Texas, USA. The instrument is a cross-dispersed white pupil echelle spectrograph.  A $2048\times 2048$ pixel CCD at the ``F3'' focal position, combined with a 1.2 arcsec wide entrance slit gives a resolving power of about 60,000 over a wavelength range of 375 to 1020 nm.  A temperature stabilized I$_2$ vapor absorption cell placed in front of the slit provides the velocity metric.  All data were reduced and extracted using an \textsc{IRAF} pipeline script.  Precise radial velocities down to a few m/s were computed using the Austral pipeline \citep{2000A&A...362..585E}.

\subsubsection{The TRES spectrograph}

We used the Tillinghast Refector Echelle Spectrograph
(TRES) on Mt. Hopkins, Arizona, to crosscheck and compare the stellar parameters with those derived from our data from the Tull, OES, and TCES spectrographs. During 2019, we obtained two spectra for each system of our targets. The spectrograph has a resolving power of $R=44 000$
and covers wavelengths from 390nm to 910nm. The relative RVs that we derive
from TRES spectra use multiple echelle orders from each
spectrum that are cross-correlated with the highest S/N
spectrum of the target star. We omit individual orders
with poor S/N and manually remove obvious cosmic
rays. 

\section{Data analysis and parameters determination}\label{sec3}

In this section, we describe the methods used to analyse our data sets to determine the stellar parameters of the host stars such as radii, masses, effective temperatures, and metallicities, as well as their ages. The photometric light curves were also used to investigate stellar rotation.

\subsection{Fitting the light curves with \textsc{TLCM}}\label{subsec:tlcm_fits}

We used for our analysis the software package \textsc{Transit and Light Curve Modeller} (TLCM, \citealt{2021arXiv210811822C}). This code is able to perform joint radial velocity and light curve fit, modelling the beaming, ellipsoidal and reflection effect \citep{2011MNRAS.415.3921F,2017PASP..129g2001S,2020AJ....159..104W}. The noise model is based on the wavelet formulation of \citet{carter_winn09}, with the regularization that the standard deviation of the residuals must converge to the average uncertainty of the input photometric data points, thereby avoiding overfitting. The approach was widely tested and the results have shown that, at the signal-to-noise ratio level of the present targets, the method is able to recover all the parameters with 1-2\% accuracy \citep{2021arXiv210811822C}. We fit the transit parameters and the noise model simultaneously. Because of the noise-model applied here, we used the SAP-light curves for this kind of analysis and we kept only those data points with a quality flag of $0$. The light curve was in the first step corrected for any potential contamination before modeling, using the {\sc crowdsap} keyword of the TESS {\sc fits}-files.

The free parameters were: the scaled semi-major axis, the planet-to-star radius ratio, the impact parameter, and the limb darkening combinations of the quadratic limb darkening law as follows:
\begin{equation}
u_+ = u_a + u_b,
u_- = u_a - u_b,
\end{equation}
as well as the epoch, period, flux zero-level shift, wavelet-based noise model $\sigma_w$, $\sigma_r$, the systematic velocity of the system ($V_\gamma$), radial velocity amplitude ($K$), and the radial velocity offsets between different spectrographs. Because only Full Frame Images (FFIs) with 30 minute integration time were available in the case of TOI-1516 and TOI-2046, we applied five subexposures and numerical integration to get more precise modelled flux values (cf. Section 2.11 of \citealt{csizmadia2020}). We have run a Genetic Algorithm/Harmony Search to optimize the joint light curve and radial velocity fit, with the results refined by three chains of Simulated Annealing.
This yielded a starting point of the MCMC-analysis. We have run four chains for each system with a thinning factor of five. The chain lengths were at least 100,000 but they were automatically extended if they did not converge. The convergence was checked via the Gelman-Rubin statistic as well as via the estimated sample size. For details, see \citet{csizmadia2020} and references therein.

As a sanity check, we performed an independent joint analysis of the
transit light curves and Doppler measurements of TOI-1181, TOI-1516, and
TOI-2046 using the code \texttt{pyaneti} \citep{2019MNRAS.482.1017B}, which infers
the system parameters using a Bayesian approach coupled to Markov chain
Monte Carlo simulations. We imposed uniform uninformative priors for all
the model parameters and added photometric and RV jitter terms to account
for noise not included in the nominal uncertainties. We initially fitted
for the eccentricity and angle of periastron adopting the 
parametrization proposed by \citet{2011ApJ...726L..19A} and found that the
eccentricities of the three planets are consistent with zero, with the
Bayesian information criterion favouring circular models. We thus fixed the
eccentricities to zero in our analysis. We found that the mean densities of the three stars -- as derived from the
modelling of the transit light curves using Kepler’s third law \citep{winn10}
-- are consistent with the spectroscopic densities. We also found that
system parameters agree well within $\sim$1$\sigma$ with those derived with
TLCM, corroborating our results.

\subsection{TOI-1181 b: A hot Jupiter on a 2.1 day orbit around a subgiant star}

We combine here \textit{TESS} with the ground-based \textit{CRCAO} and \textit{MUSCAT2} photometric data and $57$ spectra from OES, TCES and Tull spectrographs. The resulting RVs for all instruments used during the follow-up campaign are listed in Table~\ref{tabrvs1181}. We also used the TRES spectra.

\begin{figure*}
	\includegraphics[height=4.5cm,width=5.5cm]{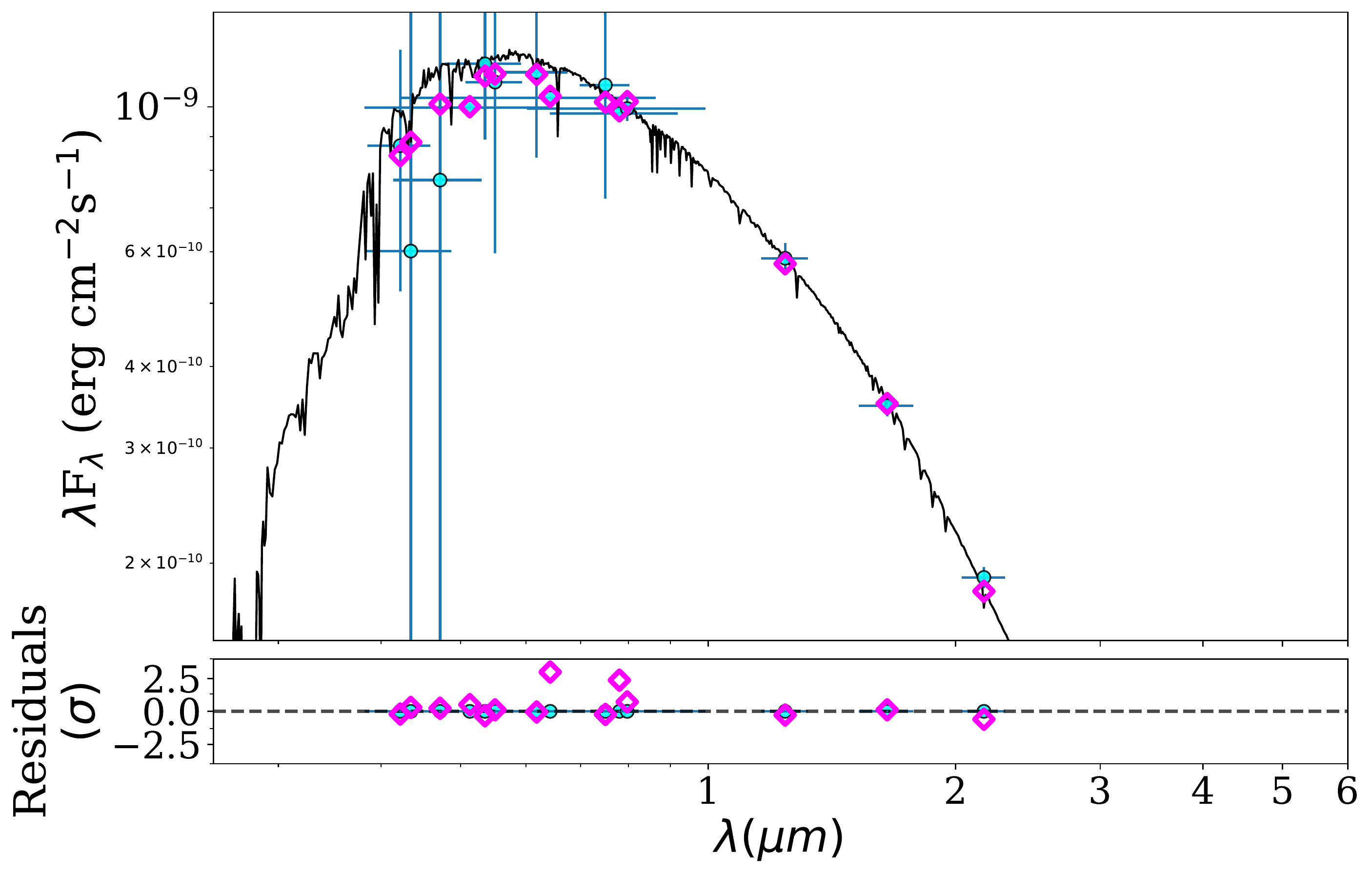}
	\includegraphics[height=4.5cm,width=5.5cm]{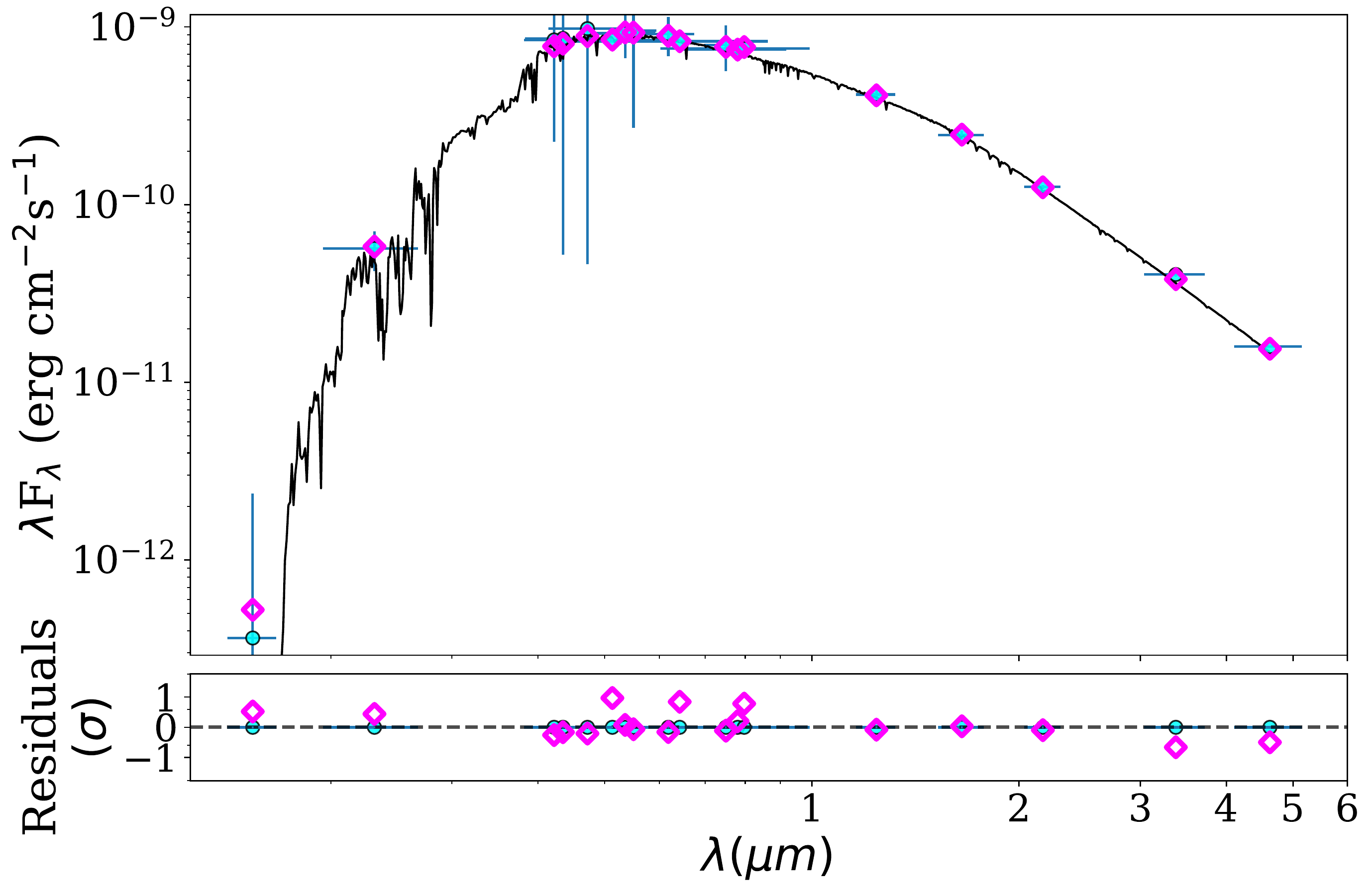}
	\includegraphics[height=4.5cm,width=5.5cm]{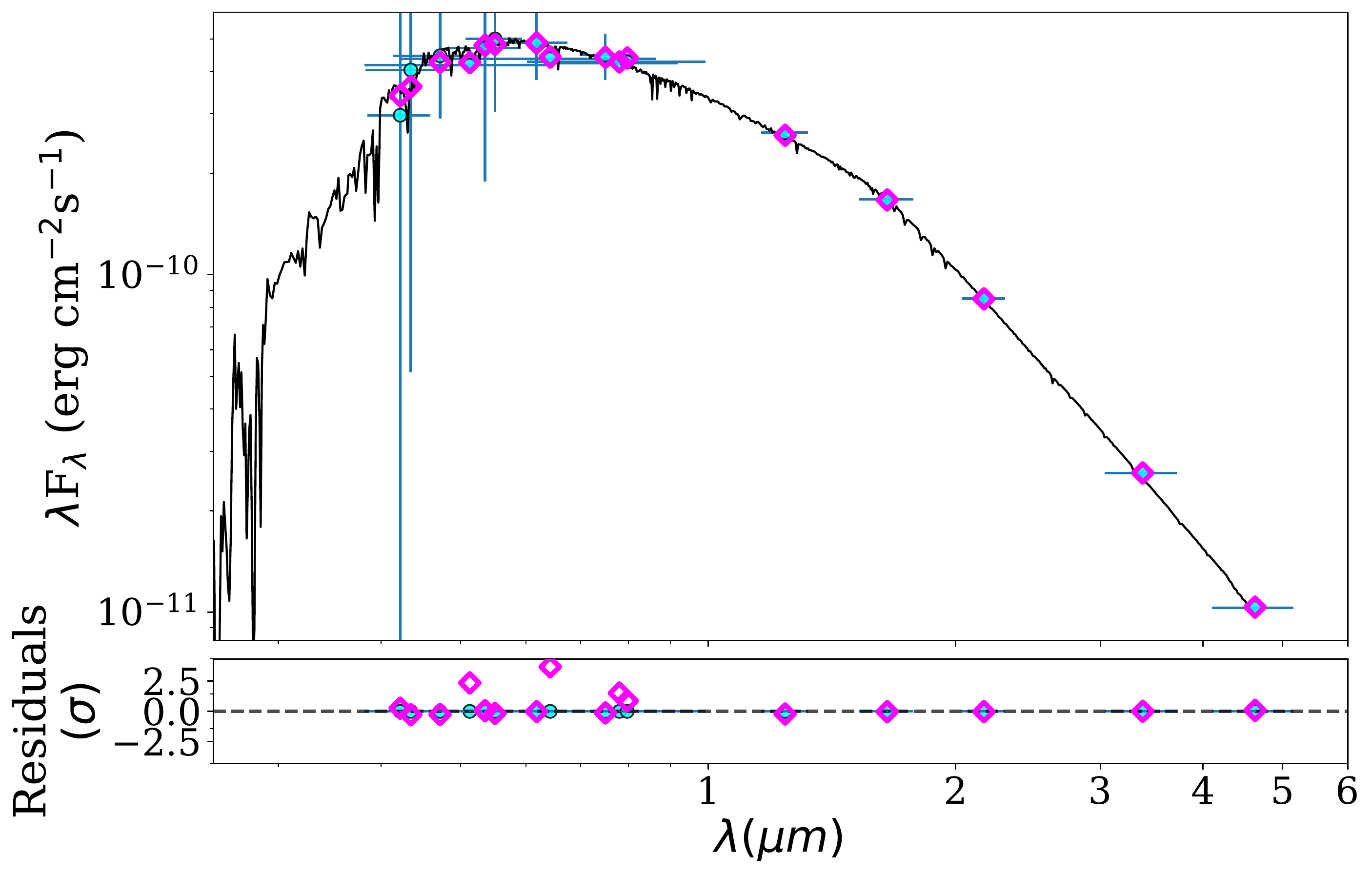}
    \caption{The SED fitted by \textsc{ARIADNE} for TOI-1181 (top), TOI-1516 (center), TOI-2046 (bottom).}
    \label{seds}
\end{figure*}

\subsubsection{The host star TOI-1181}\label{1181}

First, we combined 13 best SNR spectra from Tautenburg's TCES (i.e., only the spectra obtained without an Iodine cell and with a better SNR than the OES ones), in order to determine the stellar parameters of the host star. We used the \textsc{iSpec} software for the analysis \citep{2014A&A...569A.111B,2019MNRAS.486.2075B}. Regions between 490 and 560 nm and between 605 and 627 nm were used to derive the effective temperature, T$_{\rm eff}$, metallicity, $[{\rm Fe/H}]$, surface gravity, $\log g$ and rotational velocity of the star, $v \sin i$. We used the synthesis method from the code \textsc{SPECTRUM} and \textsc{MARCS} atmosphere models \citep{2008A&A...486..951G}. We also obtained parameters from the Tull spectra with the \textsc{Kea} pipeline \citep{2016PASP..128i4502E}, using synthetic spectra from the library compiled by \cite{1993KurCD..13.....K}. Another independent parameter set was derived from the analysis of the TRES spectra with \textsc{EXOFASTv2} \citep{2019arXiv190709480E}. The resulting values from these three independent analyses can be compared in Table~\ref{tabresu}. The same procedure applies for the other two systems discussed in this paper. It appears that the TRES and Tull values for Fe/[H] and log(g) differ from the values derived with \textsc{iSpec}. The \textsc{iSpec} values were obtained manually by inspection of the spectra, while the TRES and Tull teams use automatized pipelines which might be biased towards main sequence stellar templates, but TOI-1181 is, as described in the following, an evolved star. Therefore, for further modelling, we use the \textsc{iSpec}-derived parameters for TOI-1181, that is, effective temperature T$_{\rm eff}=5990\pm95$ K,  gravity $\log g=3.9\pm0.15$ and  $v \sin i=10.2\pm1.5$ km/s. \newline


\subsubsection{Stellar parameters}

We used the \textsc{ARIADNE}\footnote{https://github.com/jvines/astroARIADNE} code \citep{vines} to determine the Spectral Energy Distribution function (SED) and the mass of the star by fitting the available measurements from various catalogues with various models, described below. The fitted SED of TOI-1181 is shown in the top panel of Figure \ref{seds}.  \textsc{ARIADNE} uses as input value the parallax from Gaia EDR3 \citep{2018A&A...616A...1G,2016A&A...595A...1G} and models are obtained by convolution with Phoenix v2 \citep{2013A&A...553A...6H}, BT-Cond \citep{2012RSPTA.370.2765A}, BT-NextGen \citep{2012RSPTA.370.2765A,1999ApJ...512..377H}, Castelli \& Kurucz \citep{2006A&A...454..333C} and Kurucz \citep{1993KurCD..13.....K} models with the available photometric data in the following bandpasses: UBVRI; 2MASS JHKs; SDSS ugriz; ALL-WISE W1 and W2; Gaia G, RP, and BP; Pan-STARRS griwyz; Stromgren uvby; GALEX NUV and FUV; Spitzer/IRAC 3.6$\mu$m and 4.5$\mu$m; TESS; Kepler; and NGTS. We also interpolated values for $T_\mathrm{eff}$, $\log g$ and [Fe/H]. The stellar mass $M_*$ was determined from MESA Isochrones and Stellar Tracks models \citep{2016ApJS..222....8D}. The corresponding derived parameters are reported in Table \ref{tabresu}. We used as priors the values of $T_\mathrm{eff}$, $[{\rm Fe/H}]$ and $\log g$ determined from the spectra with \textsc{iSpec}. 
\textsc{ARIADNE} derived for TOI\textendash1181 a stellar mass of $M_{\ast} = 1.38^{+0.086}_{–0.082} M_{\sun}$, a radius of
$R_{\ast} = 1.9^{+0.06}_{–0.10} R_{\sun}$ and an age of $2.59^{+0.43}_{–0.51}$ Gyr.

We further derived the spectral type and luminosity class of TOI-1181 following the method outlined in \cite{2003A&A...405..149F} and
\cite{2008ApJ...687.1303G}. Briefly, we fitted the iodine-free Tull spectra of
the three stars using the Indo-US grid of stellar templates from \cite{2004ApJS..152..251V}. Prior to the fitting procedure, the resolution of the Tull
spectra was degraded to match that of the templates (R=5,500) by convolving
the Tull spectra with a Gaussian function mimicking the difference between
the two instrument profiles. The radial velocity shift between the Tull
spectra and Indo-US templates was measured using the cross-correlation
technique. We fitted the 4500–6800 \AA\ spectral range, masking out the
regions containing telluric lines. We found that TOI-1181 is a somewhat
evolved F9 IV star. Further discussion on TOI-1181 is provided in Sec \ref{sec4}.\newline  

\begin{figure}
	\includegraphics[width=\columnwidth]{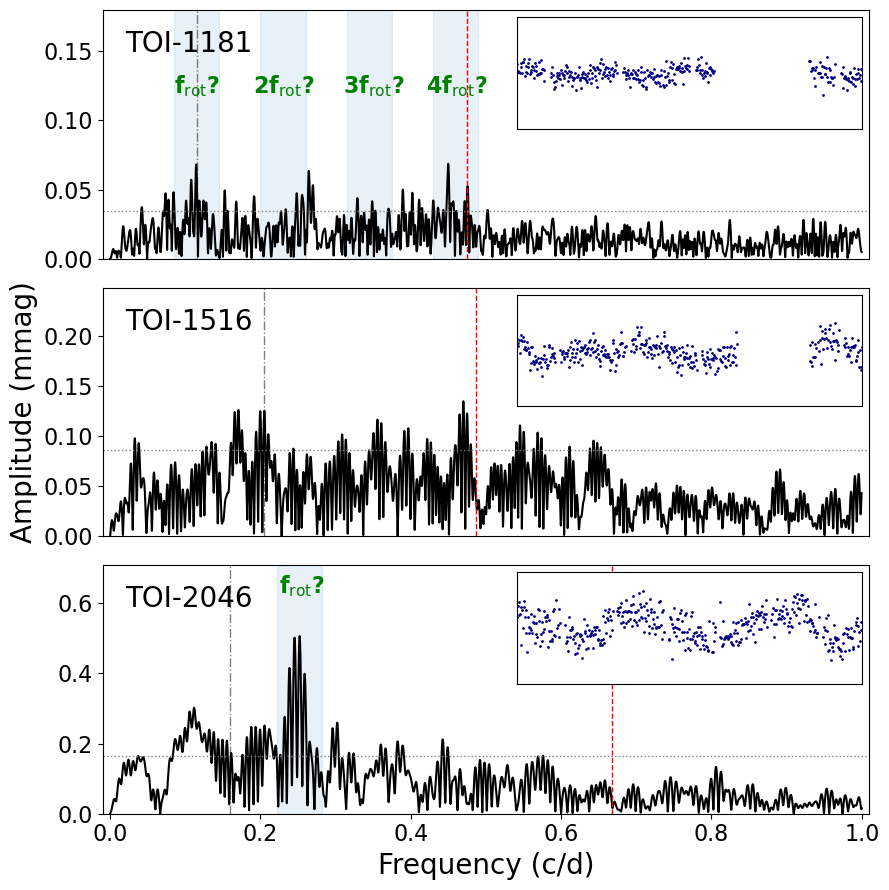}
    \caption{The frequency spectra of the residual light curves (without transits) for the
investigated stars. The red dashed vertical line shows the position of the orbital
frequency, the grey dash-dotted line shows the position of the expected rotational period calculated from $v\sin i$ and the stellar radius, while the grey dotted lines show the reliability SNR$=4$ level for the highest peak. The insets show the 10-days segments of the light curves. The vertical range of the insets is 6 mmag. }    \label{periodog}
\end{figure}

\subsubsection{Stellar rotation}

To investigate the light variations out of eclipse and their possible connection with  stellar variability, we used the PDCSAP data with 1800s cadence\footnote{Downloaded using \textsc{lightkurve} v 2.0 package \citep{2018ascl.soft12013L} because the SAP data suffer from strong instrumental effects (see the right-hand panels of Fig.~\ref{fig1}).}. After we removed the transits from the light curves, we performed a discrete Fourier transform on the full data sets by using the \textsc{Period04} software \citep[][]{Lenz2005} to get the periodograms (Fig.~\ref{periodog}). The frequency spectra of all the three studied stars show peaks above the significance limit defined by {\it SNR}$>4$ \citep{1993A&A...271..482B}. To test the nature and stability of the identified peaks, we divided the data sets into segments spanning 20-30 days and performed the frequency analysis on these shorter data sets. 

In TOI-1181, we identified three significant peaks at 0.449, 0.264 and 0.114\,c/d (periods of 2.22, 3.78 and 8.74\,days -- top panel of
Figure  \ref{periodog}). The peak corresponding to the period of 8.74\,days perfectly matches the rotational period calculated from $v\sin i$ (assuming $i=87$\,deg) and stellar radius ($R_{*}=1.75$\,R$_{\odot}$), $P_{\rm rot}=8.7\pm 1.4$\,days (the dash-dotted line in the top panel of Fig.~\ref{periodog}). We detected this frequency and its harmonics also in all the subsets, therefore making unlikely the instrumental origin of this frequency peak. It is highly likely that these peaks are real and attributed to the rotational period of the star. The differential rotation and spots at different latitudes are most likely responsible for the slight difference of the frequency peaks of the harmonics.

\subsubsection{Stellar pulsations}\label{pul}

We also performed a seismic analysis to look for solar-like oscillations using the
TESS lightcurves. To do so we used two types of lightcurves: the PDCSAP lightcurves
and lightcurves obtained with a customized aperture larger than the previous one and
that is optimized for asteroseismology. For that second dataset, we selected pixels
in the target pixel files where the average flux is above 100 e$^-$/s
and making sure that we do not pick pixels contaminated by a nearby star.
For both lightcurves we removed the planetary transits, outliers, and filled the gaps
following \cite{2011MNRAS.414L...6G,2014A&A...568A..10G}. 
Using global seismic scaling relations
\citep{1991ApJ...371..396B,1995A&A...293...87K} where we combined the
effective temperature and the surface gravity, we estimated the predicted frequency
of maximum power, $\nu_{\max}$, to be around 1000\,$\mu$Hz. From the visual inspection of the
power spectrum density (PSD) computed for both lightcurves, no obvious acoustic-mode
bump is visible around that value. We ran the A2Z pipeline
\citep{2010A&A...511A..46M} to look for the signature of the modes, without
success.

We also computed the predicted amplitude of the modes according to the seismic scaling
relations taking Teff=6000K and the Gaia luminosity of the star. We found a
predicted amplitude of 5-6\,ppm. In Fig.~\ref{lper}, we show that
the amplitude spectrum between 500 and 200\,$\mu$Hz is completely flat, dominated
by the noise with a mean value of 6.6\,ppm which is higher than the expected
amplitude of the modes. If the star was more evolved (late subgiant or early red
giant), the modes would have been at lower frequency with higher amplitudes and we
would have detected them.  Thus we can only conclude that the
star could either be on the main sequence or an early subgiant, which is consistent
with our age determination. The graphs for our analysis are presented in Appendix \ref{lper}.

\begin{figure}
	\includegraphics[width=\columnwidth]{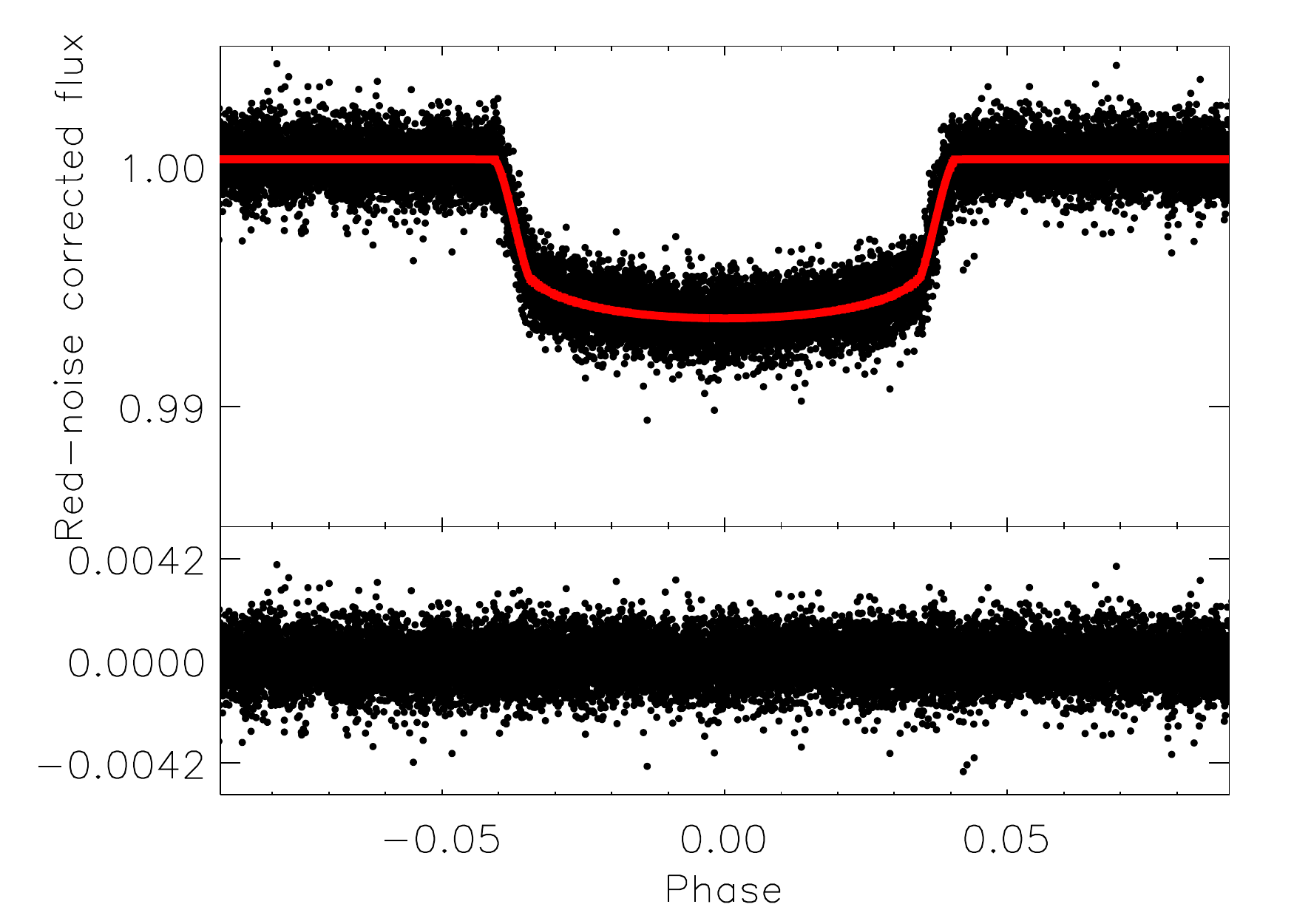}
    \caption{Upper panel: the red-noise corrected \textit{TESS} SAP-fluxes (black dots $=$ observations - red noise component) of TOI-1181 as a function of phase. The model fit is shown with red solid lines. Results obtained by TLCM (see Section~\ref{subsec:tlcm_fits}). Lower panel: the residuals of the model fit after subtracting the red noise component from the observations.}
    \label{fig:toi_1181_transitt_zoom}
\end{figure}

\begin{figure}
	\includegraphics[width=\columnwidth]{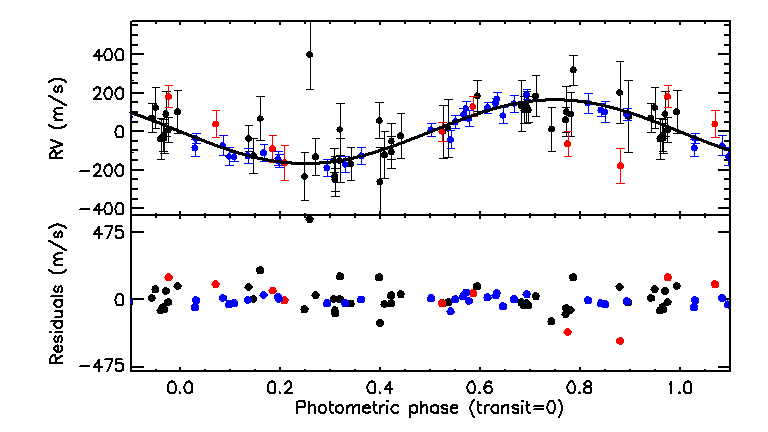}
    \caption{Upper panel: phase-folded radial velocity measurements of TOI-1181. The black dots are the TCES measurements, the red dots the OES measurements, while the blue ones represent RVs obtained by the Tull spectrograph. The black solid line shows the circular orbit fit. Lower panel: residuals of the radial velocity fit (in m/s).}
    \label{fig:toi_1181_rv}
\end{figure}

\subsubsection{TOI-1181b is a hot Jupiter around a subgiant}

As explained above, in order to derive the planetary parameters and the orbital solution, we performed a joint fit with the \textsc{TLCM} package that combined all data sets from the spectroscopic instruments and from \textit{TESS}.  The fitted ratio of $R_\mathrm{planet}/R_{*}$ and inclination $i=87.0^{\circ}{\pm 1.3^{\circ}}$ combined with the stellar radius $R_*$ provides a planetary radius $R_\mathrm{planet}=1.30\pm0.08 R_\mathrm{J}$. The planetary mass from the spectroscopic data and inclination $i$ results in $M_\mathrm{planet}=1.18\pm0.14 M_\mathrm{J}$. An orbital period $P=2.103195^{+0.000012}_{-0.000011}$ was derived from the fit. All fitted parameters, for each instrument, are presented in Table~\ref{tabresu}. The folded \textit{TESS} light curve is presented in Fig. \ref{fig:toi_1181_transitt_zoom} and RVs are presented in Fig. \ref{fig:toi_1181_rv}. Based on all our solutions, we are thus seeing a star which left the main sequence just recently and which is orbited by a hot Jupiter.

\begin{table*}
\large
	\centering
	\caption{System parameters for host stars of hot Jupiters TOI–1181, TOI–1516 and TOI–2046 obtained by TLCM. Stellar mass and radius are derived from isochrone–fitting as described in \citet{csizmadia2020}. }
	\label{tabresu}
	\begin{tabular}{lccccc} 
		\hline\hline
	Parameter              & Instrument   & TOI–1181b                   & TOI–1516b                   & TOI–2046b \\ \hline
    $a/R_{\ast}$              &   & $4.19^{+0.06}_{-0.08}$        & $6.22^{+0.041}_{-0.077}$      & $4.75^{+0.18}_{-0.17}$\\
    $R_\mathrm{planet}/R_\ast$& & $0.0764^{+0.0004}_{-0.0004}$  & $0.1224^{+0.0005}_{-0.0005}$  & $0.1213^{+0.0017}_{-0.0021}$\\
    $b$                       & & $0.19^{+0.08}_{-0.11}$        & $0.09^{+0.10}_{-0.07}$        & $0.51^{+0.06}_{-0.07}$ \\
    $u_+$                      && $0.46^{+0.06}_{-0.06}$        & $0.10^{+0.08}_{0.08}$         & $0.39^{+0.24}_{-0.16}$ \\
    $u_-$                      && $0.16^{+0.19}_{-0.19}$        & $0.81^{+0.19}_{-0.19}$        & $0.30^{+0.59}_{-0.74}$ \\
Epoch (BJD)$-2~450~000$       &&$8684.4058^{+0.0017}_{-0.0017}$&$8765.3250^{+0.0001}_{-0.0001}$ & $7792.2767^{+0.0024}_{-0.0022}$\\
    Period (days)       &&$2.103195^{+0.000012}_{-0.000011}$ & $2.056014^{+0.000002}_{-0.000002}$ & $1.4971842^{+0.0000031}_{-0.0000033}$ \\
    $\sigma_r$ (ppm)            && $34469^{+589}_{-584}$         & $14950^{+252}_{-252}$        & $24024^{+471}_{-469}$ \\ 
    $\sigma_w$ (ppm)            && $1046^{+3}_{-3}$              & $299^{+4}_{-4}$              & $576^{+14}_{-14}$ \\ 
    $V_\gamma$ (m/s)           && $119.1^{+15.7}_{-15.8}$        &  $-397.6^{39.5}_{-40.7}$     & $-12.6^{+5.5}_{-5.5}$ \\
    $K$ (m/s)                  && $165.6^{+11.4}_{-11.4}$        &  $460.7^{+9.0}_{-9.0}$       & $374.7^{+7.8}_{-7.8}$ \\\hline
    RV-offsets (m/s) & & & & \\ \hline
    Ondrejov - TCES  & & $-118.5^{+34.6}_{-34.5}$   & $-116.0.2^{+44.9}_{-43.1}$     & \\
    Tull-TCES        & & $-50477.5^{+17.1}_{-17.2}$ & $-48189.7^{+41.3}_{-40.0}$     & \\
    TCES - Tull      & &                            &                                &   $537.7^{+49.4}_{-49.4}$ \\
    Ondrejov - Tull  & &                            &                                & $-9479.6^{+43.3}_{-43.2}$\\
    \hline
    Derived parameters & & & & \\\hline
    Inclination (°)     &                 & $87.0^\circ\pm1.3^\circ$        & $90.0^\circ\pm0.4^\circ$ & $83.6^\circ\pm0.9^\circ$ \\
    Transit duration (hrs) &              & $4.057^{+0.014}_{-0.013}$       &$2.826^{+0.015}_{-0.014}$ & $2.410^{+0.032}_{-0.030}$ \\
    $\rho_\ast$ ($kg/m^3$)  &             & $312\pm16$                      & $1090\pm31$              & $890\pm98$ \\
    $M_\ast / M_\odot$       &            & $1.19\pm0.18$                   & $1.14\pm0.06$            & $1.13\pm0.19$ \\
    $R_\ast / R_\odot$        &           & $1.75\pm0.09$                   & $1.14\pm0.02$            & $1.21\pm0.07$ \\
    $M_\mathrm{planet} / M_\mathrm{J}$& & $1.18\pm0.14$                   & $3.16\pm0.12$            & $2.30\pm0.28$   \\
    $R_\mathrm{planet} / R_\mathrm{J}$& & $1.30\pm0.08$                   & $1.36\pm0.03$            & $1.44\pm0.11$ \\\hline
    Stellar parameters &&&&\\ \hline\hline
           $T_{\rm{eff}}$ [K] & Tull & $6260\pm 100$   &   $6420\pm100$ & $6160\pm100$  \\
			$T_{\rm{eff}}$ [K] & \textsc{iSpec} OES/TCES&  5990 $\pm$ 95  & 6520 $\pm$ 90  & 6250 $\pm$ 140 \\
		$T_{\rm{eff}}$ [K]	& TRES & $6177\pm50$ &      $6170\pm52$  & $6143\pm97$ \\
		log $g$ [cgs] &  Tull & $4.31\pm0.18$ & $4.25\pm0.18$ & $4.38\pm0.18$ \\
		log $g$ [cgs] &  \textsc{iSpec} OES/TCES & 3.9 $\pm$ 0.15 & 4.25 $\pm$ 0.15& 4.3 $\pm$ 0.15 \\
	    log $g$ [cgs] &  TRES & $4.27\pm0.08$ & $4.18\pm0.22$ & $4.35\pm0.08$ \\
		       Fe/H [dex] & Tull &   $0.30\pm0.12$ & $-0.14\pm0.12$ & $0.04\pm 0.12$ \\         
		      Fe/H [dex] & \textsc{iSpec} OES/TCES &  0.05 $\pm$ 0.1  & -0.05 $\pm$ 0.1 &-0.06 $\pm$ 0.15 \\
		      Fe/H [dex] & TRES &  $0.44 \pm 0.08$  & -0.05 $\pm$ 0.1 & $0.29\pm0.08$ \\
		$v_{\rm{sin}i}$ [km/s] & Tull & $10.77\pm0.30$  &$11\pm0.4$ &$8.15\pm0.15$ \\
		$v_{\rm{sin}i}$ [km/s]& \textsc{iSpec} OES/TCES & 10.2 $\pm$ 1.5  &11.8 $\pm$ 1.5  & 9.8 $\pm$ 1.6 \\
		$v_{\rm{sin}i}$ [km/s]& TRES  & $11.8\pm2.0$  & $13.8\pm2.0$  & $8.8\pm2.0$ \\\hline
    Derived stellar parameters & (\textsc{ARIADNE}) &&&\\\hline \hline
		Stellar radius - $R_\ast$ ($R_{\sun}$)& & $1.90^{+0.06}_{-0.10}$ &  $1.245^{+0.031}_{-	0.032}$ & $1.237^{+0.036}_{-0.032}$\\
		Stellar mass - $M_\ast$ ($M_{\sun}$)& &$1.380^{0.086}_{.0.082}$& $1.085^{+0.061}_{-	0.066}$&  $1.153^{+0.10}_{-0.093}$ \\
		Age of the host star (Gyr)& & $2.59^{+0.43}_{-0.51}$ & $4.82^{+2.44}_{-1.29}$ & $0.45^{+0.43}_{-0.021}$ \\
		Spectral type & & F9IV & F8V & F8V\\ \hline
	\hline 
	\end{tabular}
\end{table*}

\subsection{TOI-1516b: A hot Jupiter with a 2.05 day period}

We used the TESS light curve and spectroscopic data from Tautenburg, Ond\v{r}ejov and McDonald observatories for a joint fit. In total we used 81 spectra from all three observatories.

\begin{figure}
	 
	\includegraphics[width=\columnwidth]{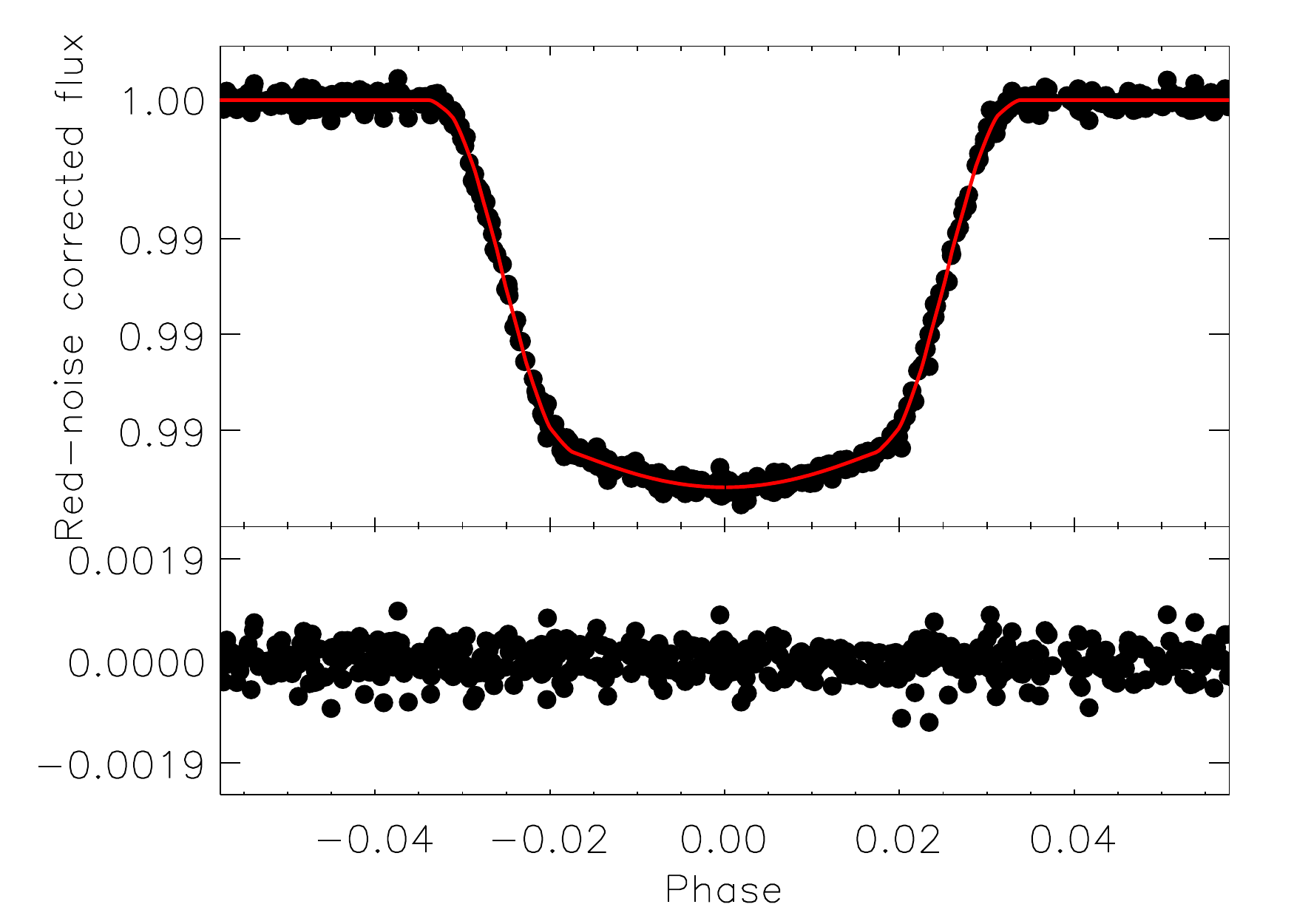}
    \caption{Same as Fig.~\ref{fig:toi_1181_transitt_zoom} for TOI-1516.}
    \label{fig:toi_1516_transitt_zoom}
\end{figure}

\begin{figure}
	\includegraphics[width=\columnwidth]{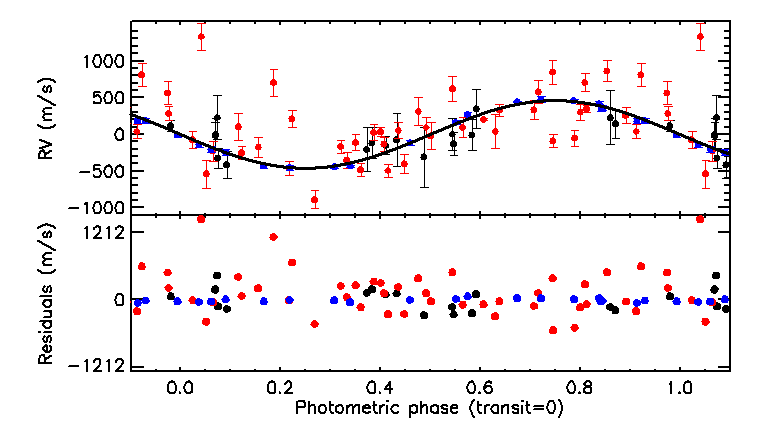}
    \caption{
    Same as Fig.~\ref{fig:toi_1181_rv} for TOI-1516.}
    \label{fig:toi_1516_rv}
\end{figure}

\subsubsection{The host Star TOI-1516}

We stacked 9 spectra from Ond\v{r}ejov in order to determine the stellar parameters of the host star. The derivation of stellar parameters with \textsc{iSpec} revealed that the host star has an effective temperature $T_{\rm eff}=6520\pm90$ K, a metallicity $[{\rm Fe/H}]=-0.05 \pm 0.1$, surface gravity, $\log g =4.25\pm0.15$ cms$^{-2}$ and $v \ sin i=11.8\pm1.5$ km/s. All parameters are presented in the corresponding central column of Table \ref{tabresu}. The spectral type and luminosity class found from our spectra as described in \ref{1181} is F8V.

A periodogram analysis was performed on the TESS data for TOI-1516. The dominant peak was found to be at 0.470\,c/d (2.13\,days), complemented with a group of other peaks in the low-frequency regime (the middle panel of Figure \ref{periodog}). We detected a peak at the expected rotational frequency calculated from $v\sin i$ and stellar radius (the grey dash-dotted line in Fig.~\ref{periodog}) also in one of the subsets. However, the detection is inconclusive.

We used \textsc{ARIADNE} to fit the SED and to determine the stellar parameters and age of the system. We used the derived stellar values from Table \ref{tabresu} as priors for \textsc{ARIADNE}. The SED obtained from the available photometric data as described in the Section \ref{1181} is depicted in the central panel of Figure \ref{seds}. The stellar mass from the isochrones was determined as $M_*=1.085^{+0.061}_{-0.066}$ $M_{\sun}$ and the stellar radius $R_*=1.245^{+0.031}_{-0.032}$ $R_{\sun}$. The age of the host star was determined as $4.82^{2.44}_{-1.29}$ Gyr. For the derivation of the above parameters, the Gaia EDR3 parallax $\varpi = 4.0540\pm0.0098$ mas was used.

\subsubsection{Planet TOI-1516b is another ordinary hot Jupiter}

We performed a joint fit with \textsc{TLCM}. We performed the first fit with eccentricity as a free parameter and based on the result being consistent with a circular orbit, we fixed the eccentricity $e$ at 0. Other parameters were left free. The resulting folded light curve from \textit{TESS} is presented in Figure \ref{fig:toi_1516_transitt_zoom} and the RVs are shown in Fig \ref{fig:toi_1516_rv}. The values of the fit confirm that TOI-1516 b is a hot Jupiter with a 2.06 day orbital period. The values for the mass and the radius are $M=3.16\pm 0.12$ $M_\mathrm{J}$ and $R=1.36\pm0.03$ $R_\mathrm{J}$, respectively. The derived parameters for the system imply an ordinary hot Jupiter around an F-type main sequence star.

\subsection{The young system TOI-2046 }

We used data from \textit{TESS} and from the Tautenburg, Ond\v{r}ejov and McDonald observatories. In total, we used 51 spectra from all three observatories. The corresponding RVs are listed in Table \ref{tabrvs}.

\begin{figure}
	 
	\includegraphics[width=\columnwidth]{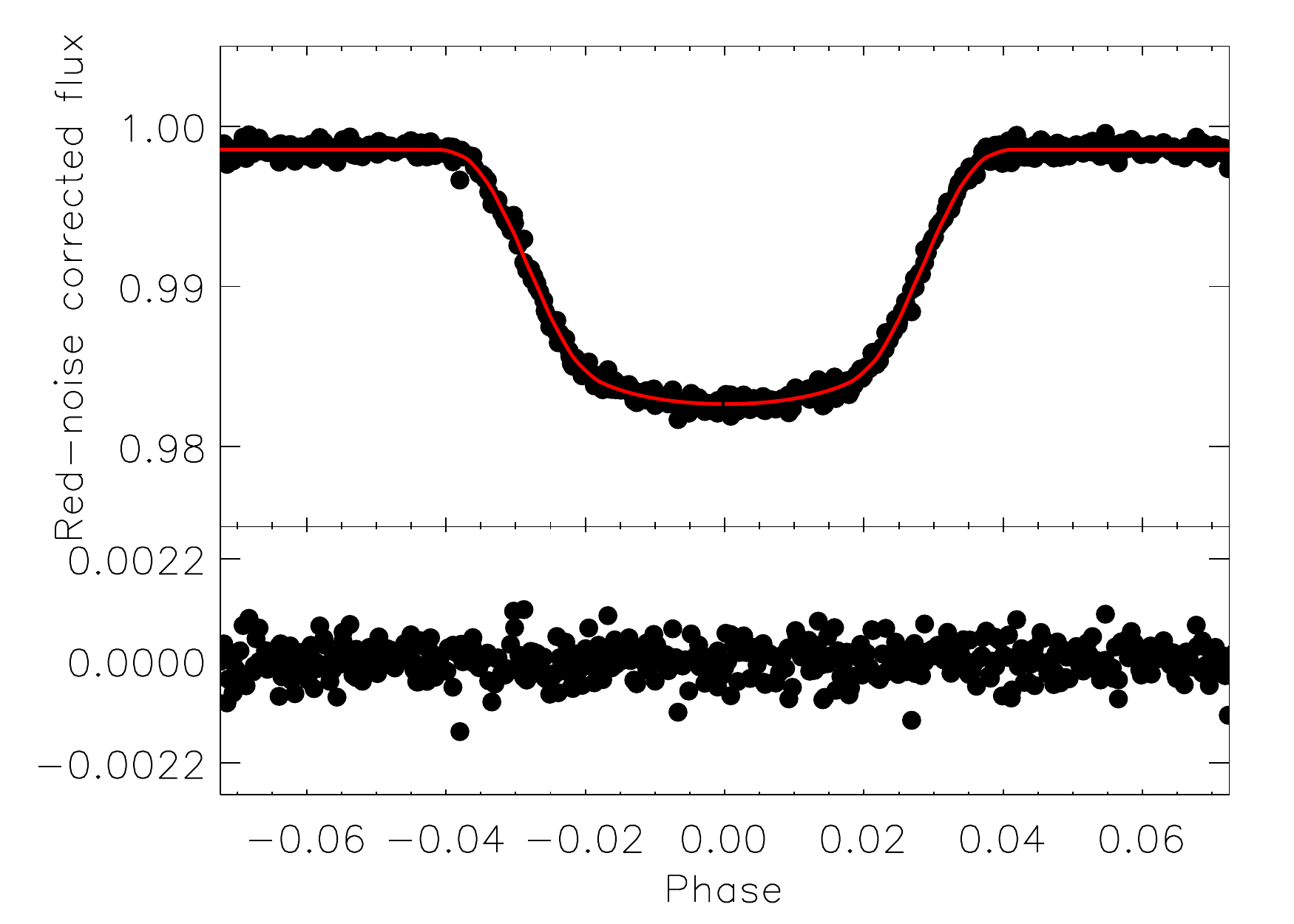}
    \caption{Same as Fig.~\ref{fig:toi_1181_transitt_zoom} for TOI-2046.}
    \label{fig:toi_2046_transitt_zoom}
\end{figure}

\begin{figure}
	 
	\includegraphics[width=\columnwidth]{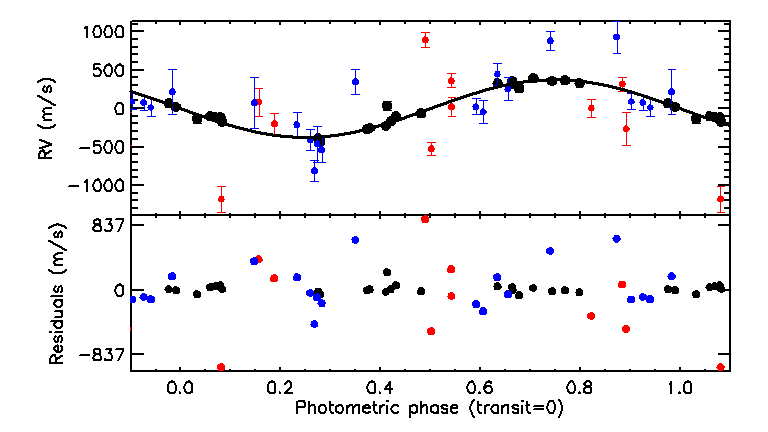}
    \caption{Same as Fig.~\ref{fig:toi_1181_rv} for TOI-2046.}
    \label{fig:toi_2046_rv}
\end{figure}

\subsubsection{The host star TOI-2046}

We stacked 7 spectra from Ond\v{r}ejov in order to determine the stellar parameters of the host star. Our analysis with \textsc{iSpec} resulted in the following parameters for the host star: $T_{\rm eff}=6250 \pm 0.140$ K, metallicity $[{\rm Fe/H}]=-0.06 \pm 0.15$, surface gravity, $\log  g =4.3\pm0.15$ cms$^{-2}$, and $v \sin i=9.8\pm1.6$ km/s. All relevant physical parameters are presented in the right column of Table \ref{tabresu}. The spectral type and luminosity class determined as described in Sec \ref{1181} is F8V.

\subsubsection{Stellar activity}

The dominant peak in the frequency spectrum of the full data set of TOI-2046 with the transit signature removed is at 0.252\,c/d (period of 3.97\,days) with a {\it SNR} $ >12$. To test the stability and reliability of the detected peak, we split the full data set into two subsets divided by the large gap between observations in sectors 18+19 (subset 1) and sectors 24 (subset 2) -- see the bottom right-hand panel of Fig~\ref{fig1}. The peak is present in both subsets and has similar amplitude (blue and red lines in the bottom right-hand panel of Fig.~\ref{periodog}). 

The flux variations with the 4-days period are clearly apparent from the inset of the bottom panel of Fig.~\ref{periodog}. The detection of the peak in the two subsets separated by more than 100 days shows that the frequency is stable over time. Thus, we assume that the variations are real (i.e., not instrumentals) and can be attributed to the stellar rotation with a period of 3.97(2)\,days. 

The value of the detected rotational period significantly differs from the period calculated from $v\sin i$ and stellar radius ($P_{\rm rot}\approx6.2$\,days). This may indicate that the rotational axis of the star and the orbital-plane axis of the planet are not aligned. The inclination of the rotational axis would be approximately 39 degrees which is significantly different from the orbital inclination of the planet (83.6 deg).

\begin{figure}
	 
	\includegraphics[width=\columnwidth]{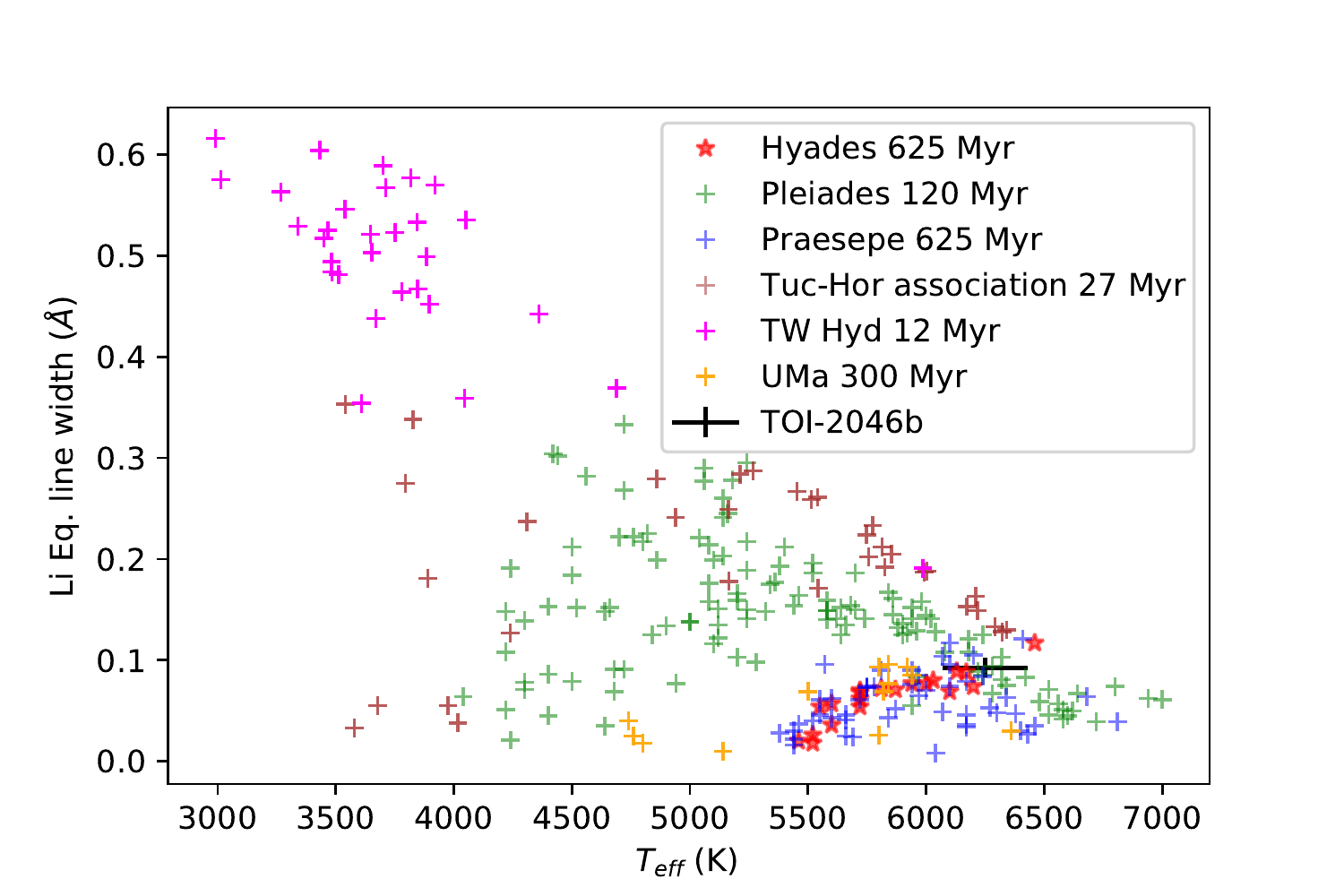}
	\includegraphics[width=\columnwidth]{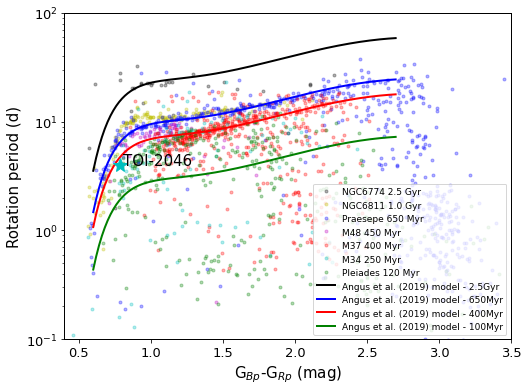}
    \caption{The top panel shows the equivalent widths of the 6708 \AA\ Li~{\sc i} line versus the effective temperatures for stars in selected young clusters and associations of different ages. The bottom panel presents the stellar rotational period versus their Gaia color index for stars in several  clusters, together with some models of cluster evolution represented with solid lines \citep{2019AJ....158..173A}. In both panels, the data  are from  \citet{1993AJ....106.1080S,1990AJ.....99..595S,2008ApJ...689.1127M,1993AJ....105.2299S} and 
    the position of TOI-2046b is indicated.}
    \label{ages}
\end{figure}

\subsubsection{Stellar age}

\textsc{ARIADNE} yields an age of $0.45^{+0.43}_{-0.02}$ Gyr for TOI-2046, indicating that this is one of the few very young known planetary systems. We tried to confirm the young age of the star using the Li line, which can be used as an indication of stellar youth. We measured the equivalent widths (EW) of a Gaussian fit to the Li line at 6707 \AA: EW$=0.083\pm0.01$ \AA. The EW and the derived stellar temperature from Table \ref{tabresu} were compared with available data from open clusters of various ages, shown in Figure \ref{ages}. A further discussion of the age of this particular system is provided in Sec. \ref{sec4} \subsubsection{A young hot Jupiter TOI-2046b}

The folded \textit{TESS} light curve is presented in Figure \ref{fig:toi_2046_transitt_zoom} and the RVs in Figure \ref{fig:toi_2046_rv}.  The parameters of the planet from the joint fit with the \textsc{TLCM} package confirm that TOI-2046b is a hot Jupiter on an orbit with a 1.49 day period and a combination with stellar parameters allows for the determination of mass $M=2.30\pm 0.28$ M$_{\rm J}$ and radius $R=1.44\pm 0.11$ R$_{\rm J}$. 
\section{New gas planets in context with other giant planets}\label{sec4}

In this section we will discuss the parameters of the newly discovered hot Jupiters in the context of similar systems. 

\subsection{Mass-Radius diagram}

\begin{figure}
	 
	\includegraphics[width=\columnwidth]{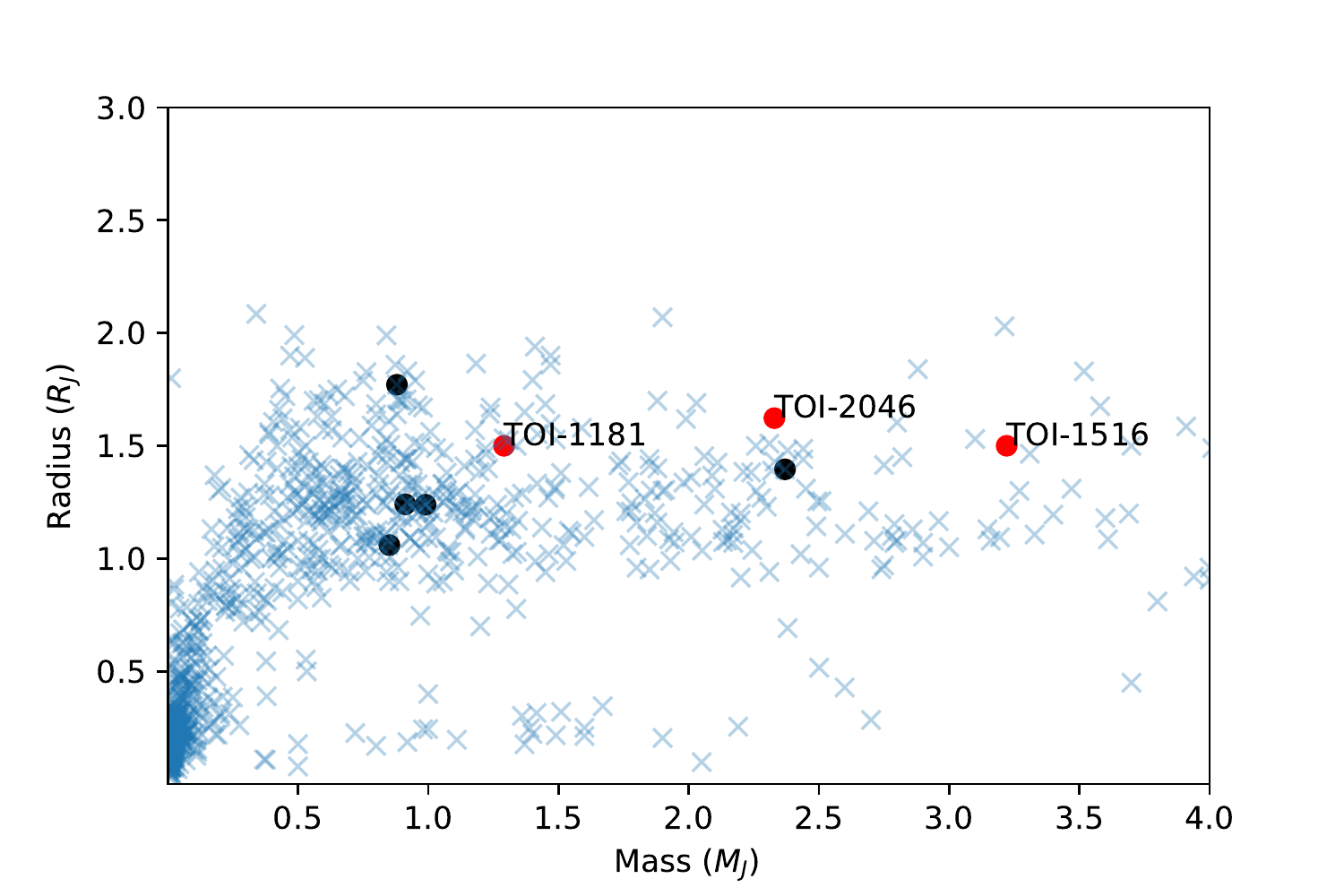}
    \caption{Mass-radius diagram of all known exoplanets with our three new systems represented by red dots. The black dots are some other recently discovered \textit{TESS} gas giants \citep{2021AJ....161..194R}, shown for comparison.}
    \label{mr}
\end{figure}

The natural metrics for comparison  of planets is the mass-radius diagram. We present the mass-radius diagram with the new \textit{TESS} giant planets from this paper and plot them with all known planets which have parameters available in \textsc{The Extrasolar planets Encyclopedia}\footnote{http://exoplanet.eu/catalog}. The three hot Jupiters described in this paper exhibit rather larger radii, as can be seen in Figure \ref{mr}. We also show in the figure a set of recently discovered \textit{TESS} hot Jupiters \citep{2021AJ....161..194R} as a comparison with similar planets.

\subsection{Incident flux from the host star and exo-atmospheres}

\begin{figure}
	 
	\includegraphics[width=\columnwidth]{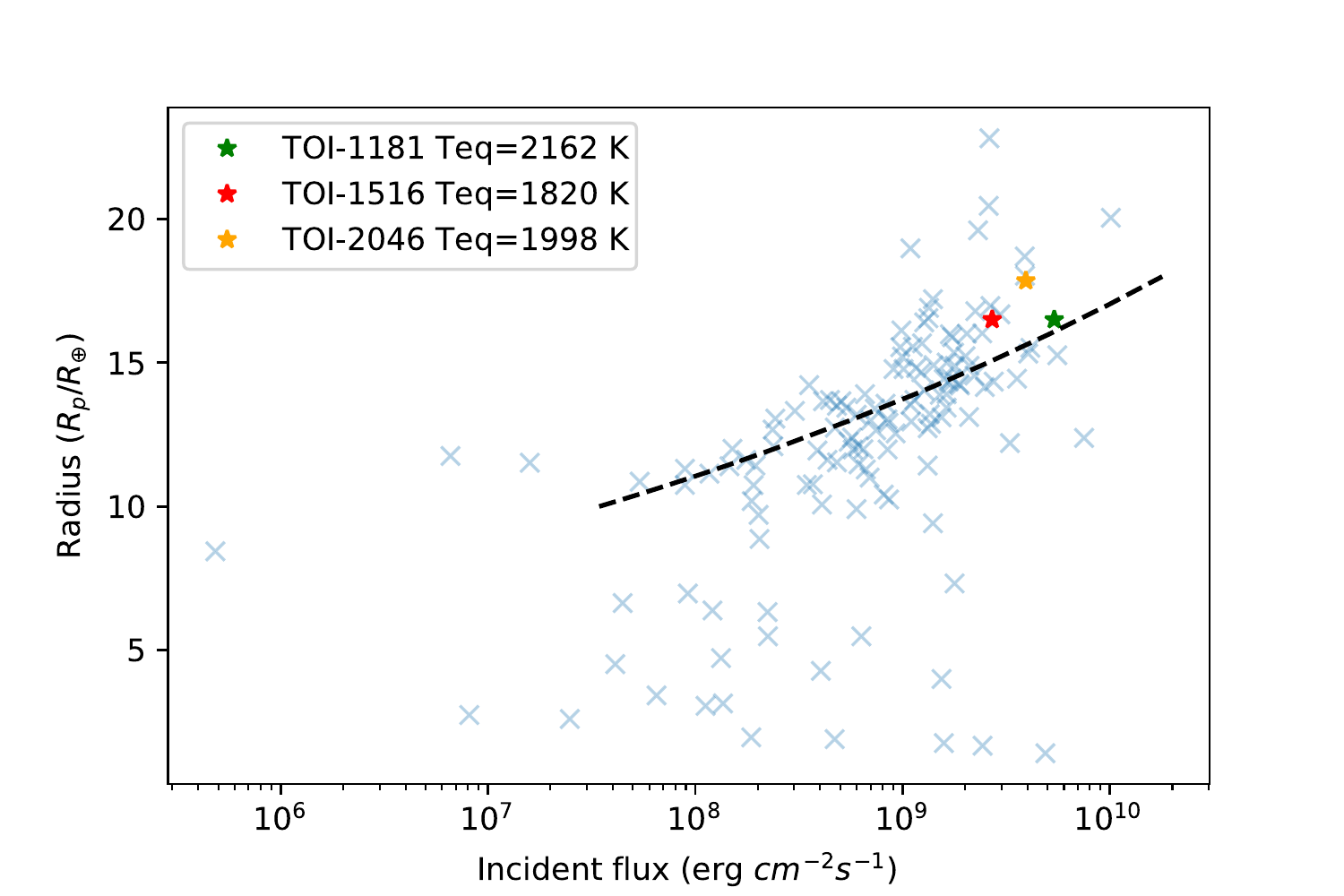}
    \caption{Radius versus incident flux for the three hot Jupiters from this paper including their equilibrium temperatures.  Other known planets are represented by blue crosses. The dashed line represents the empirical relation from \citet{2013ApJ...768...14W} for Jupiter-mass planets (Eq. \ref{flu}). }
    \label{fluxpl}
\end{figure}

We further investigated the incident flux from the host star at the planetary surface. For our three targets, we used Gaia DR2 and the luminosities determined with the \textit{FLAMES} pipeline \citep{2013A&A...559A..74B}. The resulting insolations are presented in Fig.~\ref{fluxpl}, where the planetary equilibrium temperatures are also indicated. In the same Figure, as a dashed line, we plot the empirical relationship for one Jupiter-mass gas planets derived by \citet[][their Eq. 9]{2013ApJ...768...14W}:

\begin{center}
\begin{equation}
 \frac{R_p}{R_{\earth}}= 2.45(\frac{M_p}{M_{\earth}})^{-0.039}(\frac{F}{erg.s^{-1}.cm^{-2}})^{0.094},
\label{flu}
\end{equation}
\end{center}

where $F$ is the incident flux and $R_p$ and $M_p$ are the planet's radius and mass.

\begin{figure}
	 
	\includegraphics[width=\columnwidth]{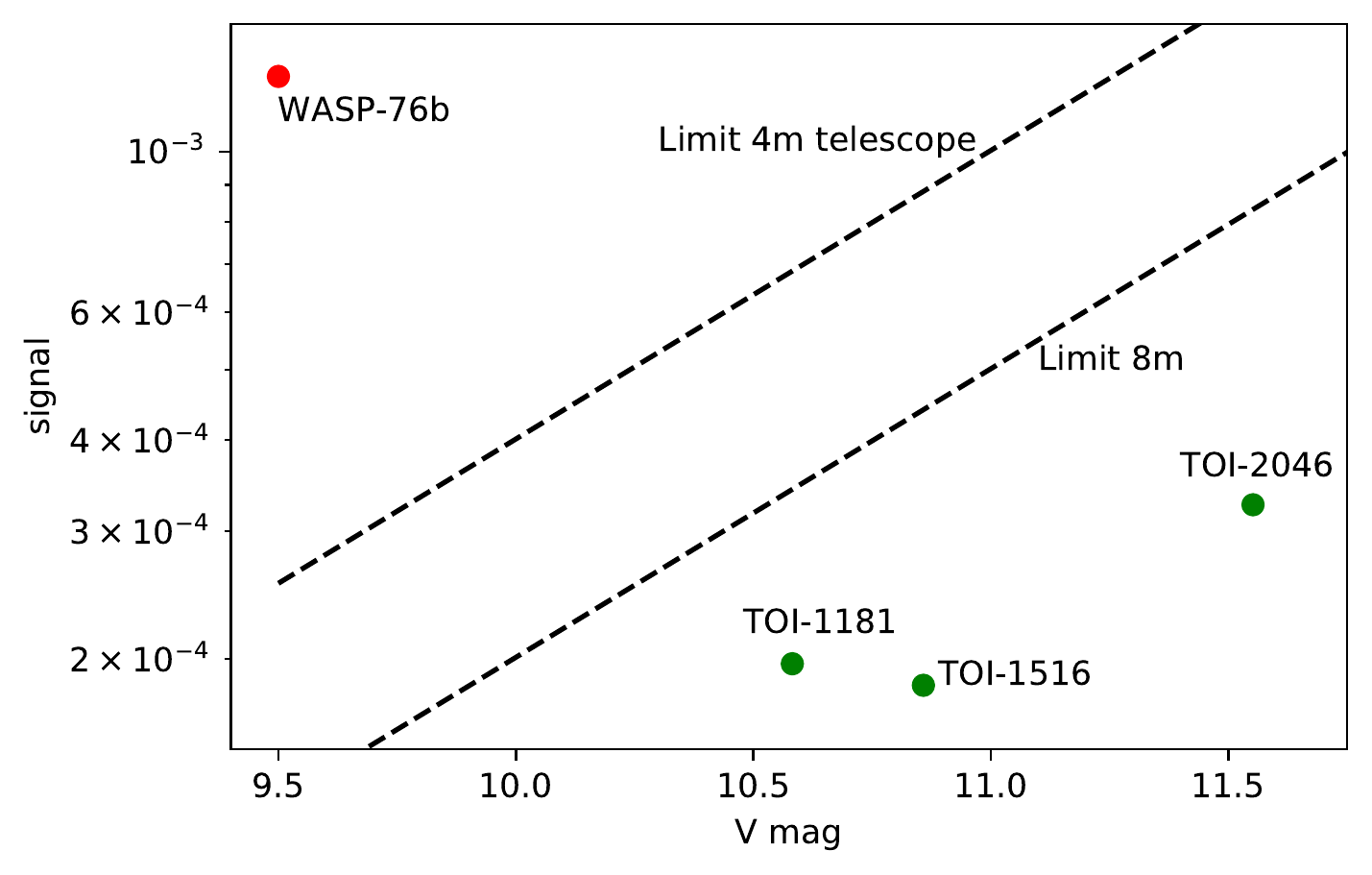}
    \caption{Expected signal of a planetary atmosphere, representative of the possible detection of the planet's NaD lines by transmission spectroscopy. We also show the position of Wasp-76b,  a reference system with a clear detection of Na, while the  dashed lines show the detection limit for two classes of telescopes.}
    \label{signal}
\end{figure}

 The three hot Jupiters reported in this paper receive  higher incident flux from their host star compared to the majority of known planets. The highest insolation is consistent with the subgiant status of TOI-1181b, but TOI-2046b also undergoes high insolation and  is most likely inflated. Therefore, these systems might in principle be reasonable candidates for further follow-up observations to detect their atmospheric signatures. For this reason, we calculated the expected atmospheric signal for performing transmission spectroscopy. We followed the approach from \citet{2019PASP..131h5001K} and we first determined the expected signature of exoatmosphere $\Delta \delta$ roughly given by:

\begin{center}
    \begin{equation}
        \Delta \delta \approx \frac{2nHR_p}{R^2_s},
    \end{equation}
\end{center}

where $n$ is the number of atmospheric scale heights $H$. Here we assume $n=6$ which is valid for high altitudes and typical for Na species abundant in gas giants. The expected signal for the three new hot Jupiters is presented in Figure \ref{signal}. We used as a reference a successful detection of the NaD features in the exoatmosphere of Wasp-76b and scaled down the detection limit at $3\sigma$ represented with a dashed line for a 4 m telescope and for an 8 m telescope. The three new planets have clearly an expected atmospheric signature too faint to be detected from the ground. However, the transmission signal $\Delta \delta=326$ ppm from TOI-2046b should be clearly detectable by \textit{JWST}, assuming the detection floor of 20 ppm for GJ1214b \citep{2014Natur.505...69K,2016ApJ...817...17G} and taking into account the fact that our star is about 1.5 mag fainter in the K-band. Planets TOI-1181b and TOI-1516b would be definitively more challenging targets. 

The high incident flux and relatively high atmospheric equilibrium temperatures make these targets interesting for detection of occultation of the planet by the star. Our team therefore analysed in detail the phased \textit{TESS} light curves of these candidates with respect to the reflected light from the planet and new results will be presented in a separate paper \citep{szilard}  for TOI-1181b. However, it would be interesting to search for an occultation of TOI-2046b due to its rather high insolation and inflated atmosphere. The limiting factor for \textit{TESS} is that TOI-2046 is one magnitude fainter than TOI-1181, leading to a much worse precision, and the analysis of the TOI-2046 \textit{TESS} light curve is inconclusive.

\subsection{A giant planet around a subgiant star}

\begin{figure}
	 
	\includegraphics[width=\columnwidth]{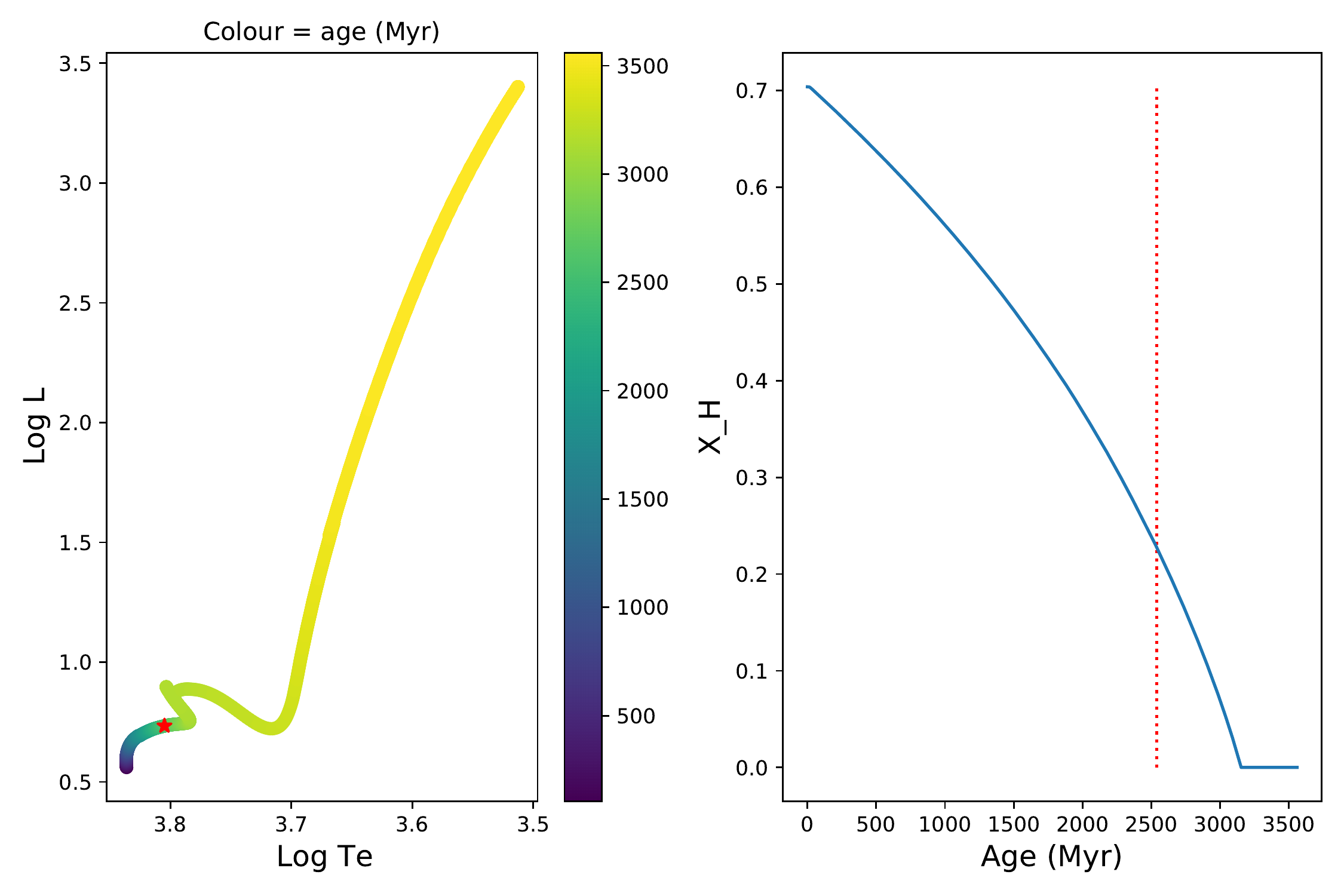}
    \caption{Left panel: PARSEC evolutionary track in the Hertzsprung-Russell diagram of a 1.38 M$_{\sun}$ star, corresponding to TOI-1181. The colours correspond to the age of the star (in Gyr), and the current position of TOI-1181 is indicated with the red dot. Right panel: Fraction of remaining hydrogen in the core of the star, based on the PARSEC track, as a function of age, with the estimated age of TOI-1181 indicated with the vertical line.}
    \label{issub}
\end{figure}

We know much less planets orbiting evolved stars than planets around main sequence stars. New planetary systems around evolved stars hosting a hot Jupiter were reported, e.g., with \textit{Kepler} and \textit{K2}, but also with TCES \citep{2016AJ....152..143V,2017AJ....154..254G,2007A&A...472..649D}. Planets around evolved stars usually show slightly larger eccentricities compared to planets around main sequence stars -- the median for hot Jupiters main sequence stars is  ${0.056}_{-0.006}^{+0.022}$, while for hot Jupiters around evolved stars it is ${0.152}_{-0.042}^{+0.077}$ \citep{2018ApJ...861L...5G}. However, in our case, we found the eccentricity of TOI-1181b to be consistent with zero. This system is also interesting due to its age of $2.59^{+0.43}_{-0.51}$ Gyr which is around the tipping point where the star leaves the main sequence and evolves towards the red giant branch, as can be seen in Fig.~\ref{issub} which shows a  PARSEC evolutionary track \citep{2012MNRAS.427..127B}. A star with a mass of 1.38 M$_{\sun}$ has left the main sequence at the age of TOI-1181, even if its core still contains about 20\% of hydrogen.

The Gaia luminosity from DR2 data corresponds to L=$4.45\pm0.08$ L$_{\sun}$. The derived $T_{\rm eff}=5990\pm95$~K corresponds to a F9 type star, but the radius $R_{\*}=1.9$ R$_{\sun}$ is too large for a main sequence star. The Gaia radius for the star TOI-1181 $R_{\ast}=2.03\pm0.07$ R$_{\sun}$ is even larger than our values and it is also indicating an evolved star. The very early stage of the subgiant branch for this star is suggested also by our asteroseismologic analysis presented in Sec. \ref{pul} as we did not detect any pulsations typical for later stages of giant stars. The star rotates still relatively fast but when compared with the sample in, e.g., \cite{2003A&A...405..723D}, its rotation is consistent with a subgiant branch stars as can be seen in their Fig. 1. Therefore, tidal locking might play a role in a slower braking of the stellar rotation. From the presented analysis and from our spectral analysis described earlier, it can be concluded that we are seeing here an early subgiant star which entered its inflationary phase.
\subsection{A very young hot-Jupiter system around TOI-2046}

We now turn to the age of the TOI-2046 system and put it in context of other similar systems. Figure \ref{ages} implies that the system TOI-2046 is a relatively young system, with an age between 27 Myr (Tucanae-Horologium association, \citealt{2008ApJ...689.1127M}) and 625 Myr as a very conservative estimate \citep[Praesepe or Hyades,][]{1993AJ....106.1080S,1990AJ.....99..595S}. We also investigated the Gaia color indices and rotational periods of selected young clusters and compared them to the TOI-2046 system's parameters (see bottom of Figure \ref{ages}) as well as to evolutionary models of young clusters from \citet{2019AJ....158..173A}. The comparison of stellar rotational period constrains the age between 120-400 Myr. However, we remark that tidal spin-up would need to be taken into account if present \citep{2021arXiv210705759T}.

\begin{figure}
	 
	\includegraphics[width=\columnwidth]{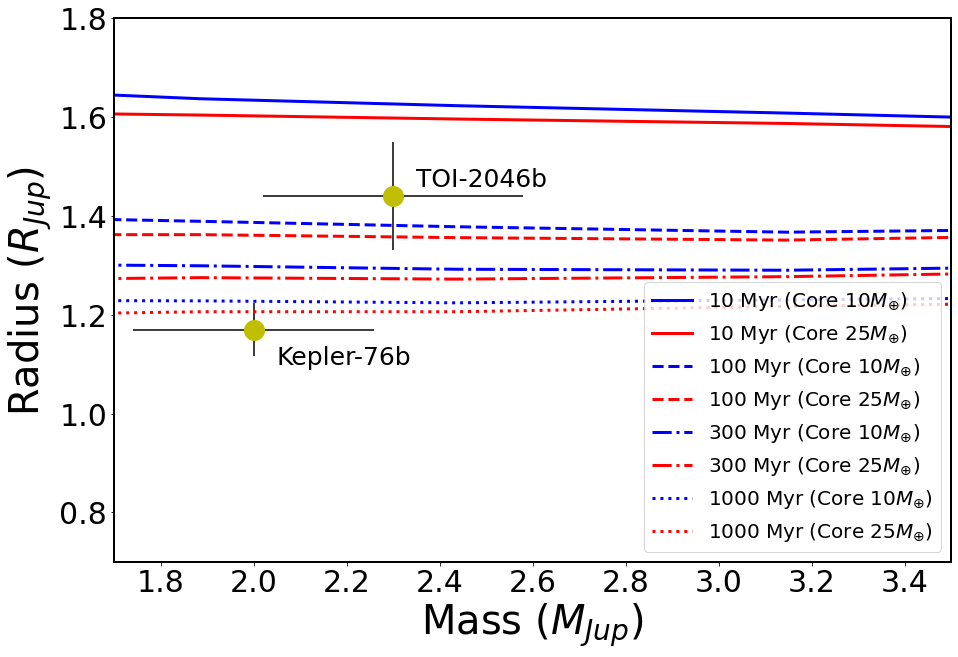}
    \caption{Mass-radius diagram with isochrones, according to thermal model by \citet{2007ApJ...659.1661F}.}
    \label{is}
\end{figure}

We used the thermal evolution model \citep{2007ApJ...659.1661F} to determine the core mass of this young system. As input parameters for the isochrones in the mass-radius graph, we use the systems parameters, the luminosity and a distance from Gaia DR2 and EDR3 release, respectively. In Fig. \ref{is}, we show the isochrones in the mass radius diagram for different core sizes and we compare the planet TOI-2046b with Kepler-76b \citep{2013ApJ...771...26F}, which has similar parameters but is older -- around 1 Gyr of age. Fig. \ref{is} would indicate an age below 100 Myr for our planet. However, it is more likely that the system is older and the planet is inflated as it undergoes high insolation. Indeed, the mechanism of inflation is not taken into account in the thermal model by \citet{2007ApJ...659.1661F} and the planet thereby appears younger in Fig. \ref{is} than it is from gyrochronology (120-400 Myr) and from comparison of the lithium content with stars in open clusters (27-625 Myr).

To date, the youngest known hot Jupiter planet is HIP 67522 b orbiting the $17\pm 2$ Myr old G1V star \citep{2020AJ....160...33R}. The second youngest gas giant orbits the $20\pm6$ Myr old 51 Eri star \citep{2015Sci...350...64M}. Another candidate for a young exoplanetary system hosting gas giants is HR 8799, but the range of age estimates is large, going from $30$ Myr to $1.1$ Gyrs \citep{2001ApJ...546..352S,2008Sci...322.1348M,2010MNRAS.405L..81M}, with a most probable solution around $155$ Myrs \citep{2012ApJ...755...38S}. Recently, a few more young systems were presented in Figure 18 of \citet{2021A&A...650A..66B}, with  ages in the range of 100-400 Myr. However, TOI-2046 is still only one of the handful fully characterized exoplanetary systems of such a very young age hosting a hot Jupiter.

Finally, we would like to stress that TOI-2046b is also a good candidate for the measurement of the Rossiter-McLaughlin effect. This has an expected amplitude of $80$ m/s. The measurement of the spin axis alignment for this young system may reveal more details of its evolutionary stage.

\section{Conclusions}\label{sec5}

In this paper we present the detection and characterization of three new hot Jupiters, TOI-1181b, TOI-1516b, and TOI-2046b, which were discovered by the \textit{TESS} space mission. We confirmed the planetary candidates by a network of instruments installed on mid-aperture sized telescopes using the \textit{KESPRINT} consortium observing time. Two of the newly detected systems, namely TOI-1181b and TOI-1516b, are hot Jupiters with short orbital periods between 1.40-2.05 days, orbiting a G type subgiant star and a main sequence F type star, respectively. TOI-2046b, orbiting a main-sequence star, is a very young planetary system with an age between 30 and 625 Myr, making it one of the handful known systems so young. Young systems are extremely important for understanding the formation and evolution of hot Jupiters and thus further studies of this planetary system, such as an analysis of the orbital dynamics, more precise determination of its age, and perhaps also potential detection of an exo-atmosphere from space would be extremely important. 

Finally, this paper clearly demonstrates that a network of telescopes with mid-sized apertures with suitable instrumentation can be used as a powerful tool for confirmation and characterization of hot Jupiter planets. Annually, about 90-135 nights of clear weather are available at Ondřejov Observatory  \citep{2020PASP..132c5002K}. Observing runs aiming for characterization of giant planets are thus ideal targets for 2-4 m instruments, leaving valuable telescope time at large observatories for smaller planets. 

\section*{Acknowledgements}

This work is done under the framework of the KESPRINT collaboration
(http://www.kesprint.science). KESPRINT is an international
consortium devoted to the characterization and research
of exoplanets discovered with space-based missions. This work is based on data sets obtained with the Perek 2-m telescope. PK, JS, MS, RK and MB are acknowledging the support by Inter-transfer grant no LTT-20015. MK acknowledges the support from ESA-PRODEX PEA4000127913. PK and JS acknowledge a travel budget from ERASMUS+ grant 2020-1-CZ01-KA203-078200. HB's mobility was funded under: MŠMT - CZ.02.2.69/0.0/0.0/18\_053/0016972 Podpora mezin\'{a}rodn\'{i} spolupr\'{a}ce v astronomii. PC acknowledges the generous support from Deutsche Forschungsgemeinschaft (DFG) of the grant CH 2636/1-1. PR acknowledges support from National Science Foundations (NSF) grant No.
1952545. This work was generously supported by the Th\"uringer
Ministerium f\"ur Wirtschaft, Wissenschaft und Digitale Gesellschaft. We acknowledge the use of public TESS data from pipelines at the TESS Science
Office and at the TESS Science Processing Operations Center. Resources supporting this work were provided by the NASA High-End Computing (HEC) Program through the NASA Advanced Supercomputing (NAS) Division at Ames Research Center for the production of the SPOC data products. JIV acknowledges support of
CONICYT-PFCHA/Doctorado Nacional-21191829. This work is partly supported by JSPS KAKENHI Grant Number JP20K14518. JK gratefully acknowledge the support of the Swedish National Space 
Agency (SNSA; DNR 2020-00104). 
HD acknowledges support from the Spanish Research Agency of the Ministry of
Science and Innovation (AEI-MICINN) under grant PID2019-107061GB-C66, DOI:
10.13039/501100011033.
R.L. acknowledges financial support from the Spanish Ministerio de 
Ciencia e Innovación, through project PID2019-109522GB-C52, and the 
Centre of Excellence "Severo Ochoa" award to the Instituto de 
Astrofísica de Andalucía (SEV-2017-0709).
This work has made use of data from the European Space Agency (ESA) mission
Gaia (\url{https://www.cosmos.esa.int/gaia}), processed by the Gaia
Data Processing and Analysis Consortium (DPAC,
\url{https://www.cosmos.esa.int/web/gaia/dpac/consortium}). Funding for the DPAC
has been provided by national institutions, in particular the institutions
participating in the Gaia Multilateral Agreement. This is University of Texas Center for Planetary Systems Habitability 
contribution \#0054.

LGC acknowledges support from grant FPI-SO from the Spanish Ministry of Economy and
Competitiveness (MINECO) (research project SEV-2015-0548-17-2 and predoctoral
contract BES-2017-082610).
S.~M. acknowledges support by the Spanish Ministry of Science and Innovation with
the Ramon y Cajal fellowship number RYC-2015-17697 and the grant number
PID2019-107187GB-I00. 
R.~A.~G. acknowledges the support from PLATO and GOLF CNES grants.

This paper includes data collected with the TESS mission, obtained from the MAST data archive at the Space Telescope Science Institute (STScI). Funding for the TESS mission is provided by the NASA Explorer Program. STScI is operated by the Association of Universities for Research in Astronomy, Inc., under NASA contract NAS 5–26555.
Observations in the paper made use of the NN-EXPLORE Exoplanet and Stellar Speckle Imager (NESSI). NESSI was funded by the NASA Exoplanet Exploration Program and the NASA Ames Research Center. NESSI was built at the Ames Research Center by Steve B. Howell, Nic Scott, Elliott P. Horch, and Emmett Quigley. The authors are honored to be permitted to conduct observations on Iolkam Du'ag (Kitt Peak), a mountain within the Tohono O'odham Nation with particular significance to the Tohono O'odham people. 
This work is partly supported by JSPS KAKENHI Grant Number JP17H04574 and
JP18H05439, JST PRESTO Grant Number JPMJPR1775, the Astrobiology Center of
National Institutes of Natural Sciences (NINS) (Grant Number AB031010).
This article is based on observations made with the MuSCAT2 instrument,
developed by ABC, at Telescopio Carlos Sánchez operated on the island of
Tenerife by the IAC in the Spanish Observatorio del Teide.

Observations in the paper made use of the High-Resolution Imaging instrument ‘Alopeke. ‘Alopeke was funded by the NASA Exoplanet Exploration Program and built at the NASA Ames Research Center by Steve B. Howell, Nic Scott, Elliott P. Horch, and Emmett Quigley. ‘Alopeke was mounted on the Gemini North telescope of the international Gemini Observatory, a program of NSF’s NOIRLab, which is managed by the Association of Universities for Research in Astronomy (AURA) under a cooperative agreement with the National Science Foundation on behalf of the Gemini partnership: the National Science Foundation (United States), National Research Council (Canada), Agencia Nacional de Investigación y Desarrollo (Chile), Ministerio de Ciencia, Tecnología e Innovación (Argentina), Ministério da Ciência, Tecnologia, Inovações e Comunicações (Brazil), and Korea Astronomy and Space Science Institute (Republic of Korea).
    
This work was enabled by observations made from the Gemini North telescope, located within the Maunakea Science Reserve and adjacent to the summit of Maunakea. We are grateful for the privilege of observing the Universe from a place that is unique in both its astronomical quality and its cultural significance. We thank the anonymous referees  for the useful suggestions to improve the paper.

\section*{Data Availability}

The article presents results based on data sets from the \textit{TESS} mission, which are available at Mikulski Archive. Furthermore, the radial velocity data values are provided in the article whereas the raw and calibrated data are available upon request to the observatory archives of Tautenburg and Ond\v{r}ejov and from the authors with an affiliation at the University of Texas for Tull data.



\typeout{}
\bibliographystyle{mnras}
\bibliography{mnras_hotjupitersaccept} 



\clearpage
\appendix

\section{Additional tables and figures}

\begin{figure*}
	\includegraphics[width=\columnwidth]{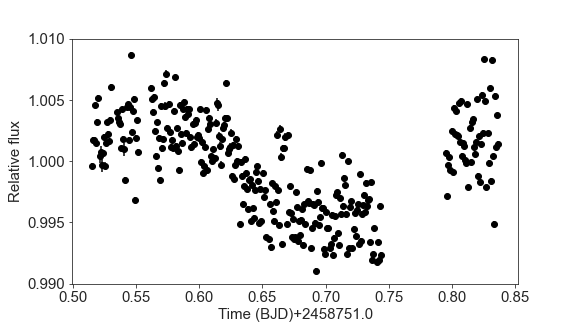}
	\includegraphics[width=\columnwidth]{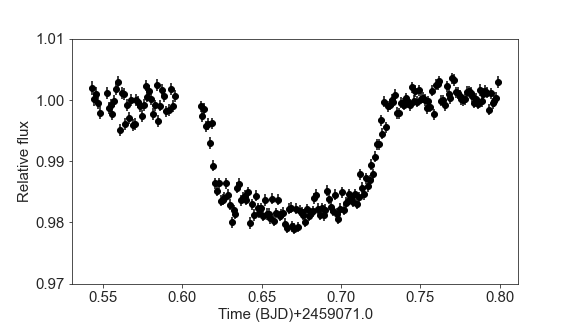}
    \caption{Light curves of TOI-1181 (left) and TOI-1516 (right) from the CRCAO Observatory.}
    \label{lcscrcao}
\end{figure*}

\begin{figure*}
	\includegraphics[height=12cm,width=16cm]{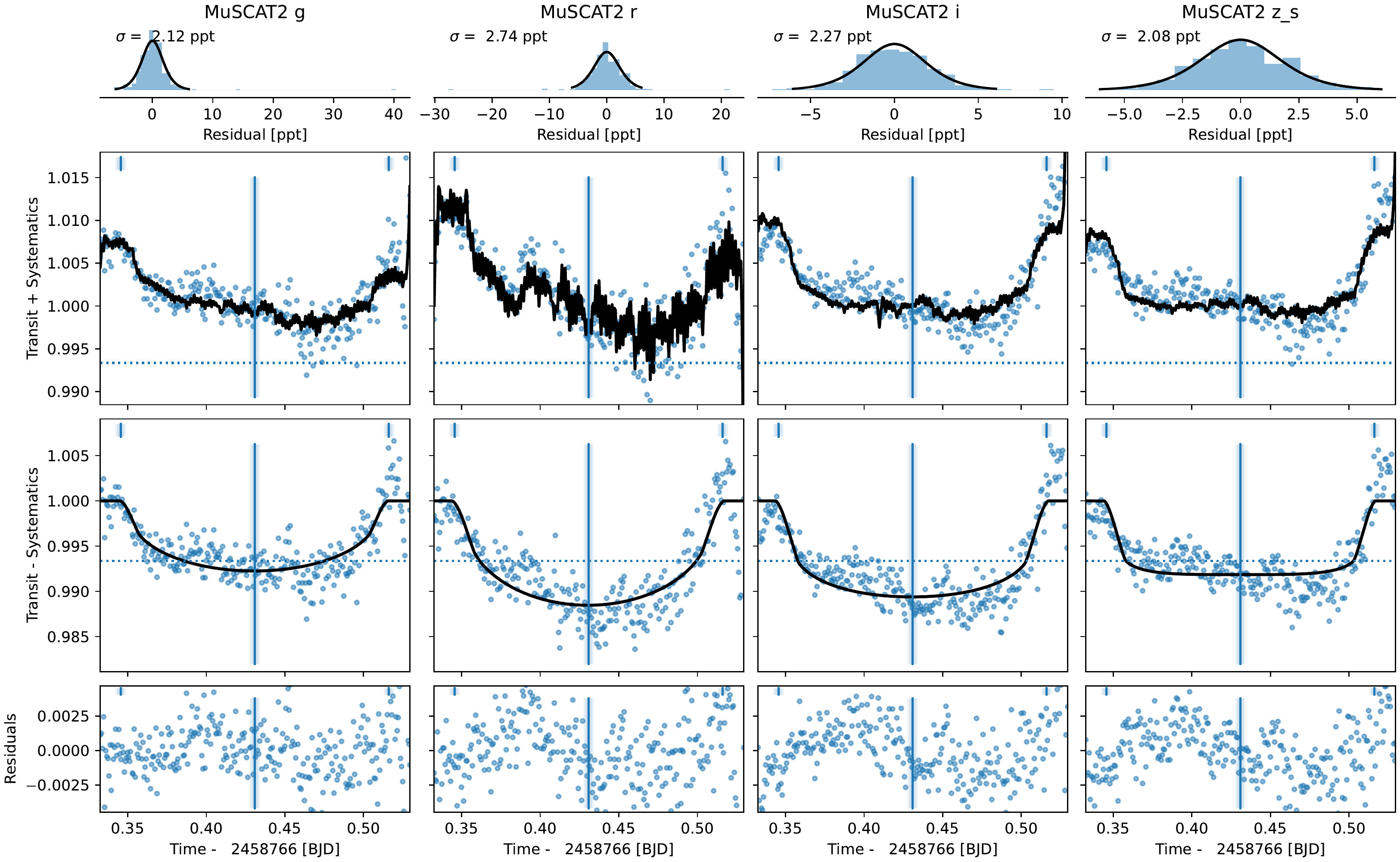}
    \caption{Light curves of the transit of TOI-1181b in different filters obtained with \textit{MUSCAT2}}.
    \label{lcmuscat}
\end{figure*}

\begin{figure*}
	\includegraphics[width=\columnwidth]{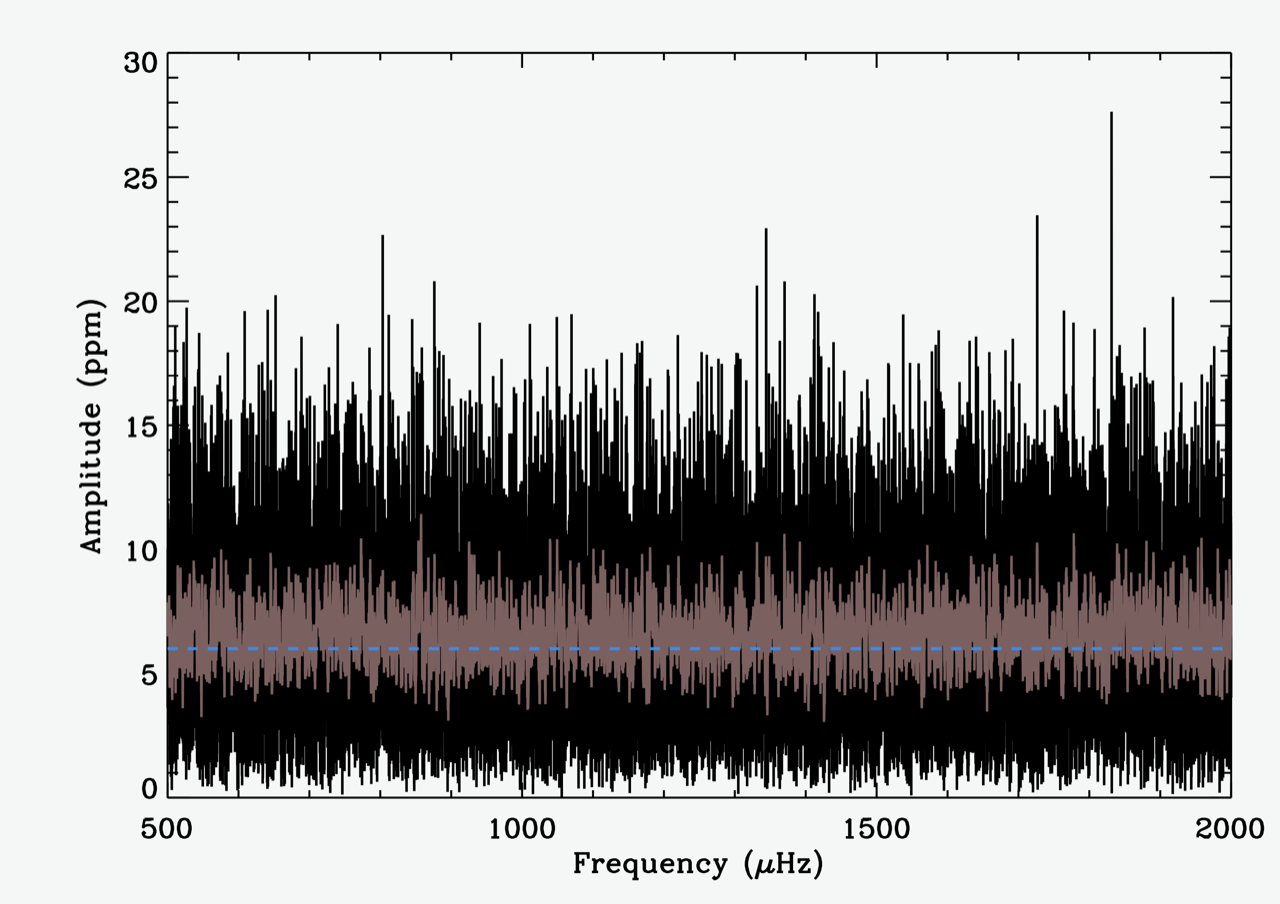}
    \caption{Amplitude spectrum as a function of the frequency (in black) where we superimposed a
smoothed curve with an 11-point boxcar (in grey). The dash blue line corresponds to
the expected amplitude of the modes of 6\,ppm.}
    \label{lper}
\end{figure*}

\begin{table*}
\caption{Parameters determined from the \textit{MUSCAT2} light curve for TOI-1181.}\label{muscat2tab}
\begin{tabular}{llll}
\hline
\hline
Epoch & 2458766.43048 &$\pm$& 0.00054\\
Period & 2.10319 &$\pm$& 0.00001 d\\
Stellar density & 0.32550 &$\pm$& 0.00787 g/cm$^3$\\
Impact parameter & b < 0.28759 && (99\% posterior upper limit)\\
Radius ratio & 0.08683 &$\pm$& 0.00132\\
\hline\hline
\end{tabular}
\end{table*}

\begin{table*}
	\centering
	\caption{Radial velocities of TOI-1181}
	\label{tabrvs1181}
	\begin{tabular}{lccr} 
		\hline
		\textbf{Date (BJD)} & \textbf{RV (m/s)} & \textbf{$\sigma_{RV} (m/s)$} & \textbf{Instrument}\\
		\hline
2458891.624504 &	119.2 &	137.2 &   TCES \\
2458891.646611 &	135.3 &	148.6 &   TCES \\
2458892.516390 &	244.2 &	106.3 &   TCES \\
2458892.538323 &	 82.2 &	107.6 &   TCES \\
2458892.560290 &	210.4 &	118.0 &   TCES \\
2458913.581137 &	 99.8 &	 84.3 &   TCES \\
2458913.602628 &	127.5 &	 65.5 &   TCES \\
2458916.594736 &	176.2 &	 94.2 &   TCES \\
2458916.616264 &	 -1.7 &	122.9 &   TCES \\
2458918.510222 & -113.1 &	104.0 &   TCES \\
2458918.532270	 &  130.4 &	136.5 &   TCES \\
2458923.503219 &	252.3 &	130.6 &   TCES \\
2458923.524388 &	246.5 &	 83.3 &   TCES \\
2459000.453694 &	-13.5 &	 96.6 &   TCES \\
2459001.510255 &	222.5 &	130.5 &   TCES \\
2459003.484689 &	303.2 &	106.1 &   TCES \\
2459006.530164 &	186.2 &	114.4 &   TCES \\
2459027.512966 &	 83.1 &	 68.2 &   TCES \\
2459066.549074 &	232.7 &	 68.3 &   TCES \\
2459067.494285 &	 -7.4 &	 91.5  &  TCES \\
2459070.540453 &	304.1 &	 80.3 &   TCES \\
2459094.407467 &	188.7 &	 76.3 &   TCES \\
2459094.516228 &	221.0 &	113.7 &   TCES \\
2459095.456436 &	 96.7 &	117.3 &   TCES \\
2459097.519515 &	 68.9 &  95.2 &   TCES \\
2459098.518832 &	196.8 &	187.3 &   TCES \\
2459099.381413 &	-31.6 &	 61.8 &   TCES \\
2459100.389551 &	438.4 &	 77.1 &   TCES \\
2459100.585164 &	322.4 &	160.9 &   TCES \\
2459101.555221 &	-49.1 &	 85.7 &   TCES \\
2459177.393867 & -142.2 &	180.3 &   TCES \\
2459177.441483	 &   13.1 &	 82.7 &   TCES \\
2459179.306668 & -130.3 &	 59.4 &   TCES \\
2459179.327816 &	-30.9 &	114.0 &   TCES \\
2459180.278832 &	180.0 &	 63.5 &   TCES \\
2459180.300000 &	208.8 &	114.9 &   TCES \\
2459181.281956 & -113.7 &	122.3 &   TCES \\
2459181.303683 &	519.4 &	179.6 &   TCES \\
2459209.548487 &	231.2 &  71.8 &   TCES \\
2459209.663867	 &	130.9 &	113.6 &   TCES \\
2459214.340952 &	 82.9 &	 79.5 &   TCES \\
2459067.574725   &	-89.0 &	 72.0 &   OES \\
2459070.521985	 &  131.0 &	 54.0 &   OES \\
2459071.541575   &   38.0 &	 70.0 &   OES \\
2459074.598145   &    0.0 &	 50.0 &   OES \\
2459075.549465   &	182.0 &  57.0 &   OES \\
2459086.554015   & -163.0 &	 91.0 &   OES \\
2459100.588335	 & -177.0 &	 91.0 &   OES \\
2459104.572505	 &  -64.0 &	 66.0 &   OES \\
2458951.925755	 & -136.7 &	 32.9 &   Tull \\
2458957.885446   &  -21.7 &	 27.8 &   Tull \\
2458994.903263   &	163.2 &	 25.0 &   Tull \\
2459054.772528   & -114.8 &	 46.2 &   Tull \\
2459072.686735   &	138.7 &	 29.8 &   Tull \\
2459073.722383   & -117.1 &	 21.6 &   Tull \\
2459090.733304   & -124.8 &	 38.8 &   Tull \\
2459091.680828   &	 98.1 &	 42.1 &   Tull \\
2459104.729921   &	115.1 &	 55.3 &   Tull \\
	
		\hline
	\end{tabular}
\end{table*}


\clearpage
\onecolumn

\begin{longtable}{lccc}
\caption{Radial velocities of TOI-1516.} \label{tabrvs} \\

\hline \multicolumn{1}{c}{\textbf{Date (BJD)}} & \multicolumn{1}{c}{\textbf{RV (m/s)}} & \multicolumn{1}{c}{\textbf{$\sigma_{RV}$ (m/s)}} & 
\multicolumn{1}{c}{\textbf{Instrument}} \\ \hline 
\endfirsthead

\multicolumn{4}{c}%
{{\bfseries \tablename\ \thetable{} -- continued from previous page}} \\
\hline \multicolumn{1}{c}{\textbf{Date (BJD)}} & \multicolumn{1}{c}{\textbf{RV (m/s)}} & \multicolumn{1}{c}{\textbf{$\sigma_{RV}$ (m/s)}} & \multicolumn{1}{c}{\textbf{Instrument}} \\ \hline 
\endhead

\hline \multicolumn{4}{r}{{Continued on next page}} \\ 
\endfoot

\hline 
\endlastfoot

2459209.57482 &	-177 &	302  &   TCES \\
2459209.61444 &	-821 &	179 &    TCES \\
2459214.38049 &	-556 &	127  &  TCES \\
2459094.43088 &	-408 &	214 &   TCES \\
2459095.50450 &	-56 &	268  &  TCES \\
2459097.54374 &	-417 &	209 &   TCES \\
2459098.54208 &	-422 &	182 &   TCES \\
2459099.40184 &	-714 &	408 &   TCES \\
2459099.51834 &	-405 &	221 &   TCES \\
2459100.41467 &	-291 &	81 &   TCES \\
2459100.60848 &	-730 &	137 &   TCES \\
2459101.57987 &	-534 &	152 &   TCES \\
2459177.41747 &	-483 &	363 &   TCES \\
2459179.35048 &	-611 &	300 &   TCES \\
2459179.37221 &	-524 &	213  &   TCES \\
2459180.35372 &	-178 &	366 &   TCES \\
2459180.37488 &	-264 &	190  &   TCES \\
2459209.57482 &	-177  &	302  &   TCES \\
2459209.61444 &	-821 &179 &   TCES \\
2459214.38049 &	-556 &	127 &   TCES \\
2458863.25393    &	    -474	 &    233 & OES \\
2458864.25126	 &  -409     &    180 & OES \\
2458865.26129	 &  -310	 &    20  &   OES \\
2458872.34369   &   -1054	& 191 	 &      OES \\
2458892.27580	& -604	  & 90  	 &      OES \\
2458892.36445	& -568	  &   110 	&  OES \\
2458892.40691	 &  193	  &  123 	&  OES \\
2458892.63805	& 298	  &     152 &     OES \\
2458898.36529	 & -179   &	130 	 &  OES \\
2458901.27357	 & -764	  & 92  	 & OES \\
2458924.32249	 & -867	  & 105 	 & OES \\
2458924.55897	 & -915	  & 128 	 & OES \\
2458925.31019	 & -165	  & 43  	 & OES \\
2458926.51850	 & -478	  & 60       &OES \\
2458928.30224	 & -1404  & 124      & OES \\
2458928.47061	 &  -626 &   108  &  OES \\
2458931.27899	 &  62	  & 155  & OES \\
2458931.56240	 & 347	  & 146  & OES \\
2458932.30872	&-973	  & 96  &  OES \\
2458932.60401	&-1000	  &  87  & OES \\
2458936.29387	&-690	  &   138  & OES \\
2458936.63373	& -679	  & 88   & OES \\
2458937.28739	& -184	  & 81  & OES \\
2458937.62028	& -213	  & 104  &  OES \\
2458942.60106	&  -302	 &115   & OES \\
2458946.38789	& -745	  &137   &  OES \\
2458946.63689   & 191	  & 183   &  OES \\
2458956.31205	& -258	  &   109  &  OES \\
2458956.35458	&  -475  & 95     & OES \\
2458956.58365	& -588	 & 116     & OES \\
2458956.61892	&  817	  & 183 &   OES \\
2458957.32975	& -489	 & 113   &  OES \\
2458957.37191	& -649	 &  97  &  OES \\
2458959.56920	&-201	 & 190  & OES \\
2458960.59308	& 47	  & 154   &  OES \\
2458963.55698	& -1007	  & 90    & OES \\
2458963.59917	&  -461	  &  111  &  OES \\
2458976.57060	& 331	  & 163   &  OES \\
2458990.55013	& 104	  &  167  &  OES \\
2458992.51773	&-536	  & 187    & OES \\
2459101.61717	& -419	   & 141   & OES \\
2459105.57857	&  -419	      & 139 &  OES \\
2459106.57426	&  -229	      &95   &OES \\
2458824.63161  & -48227    & 35  & Tull  \\ 
2458824.77300 & -48406     &34 &   Tull \\
2459054.89122  & -48166   &  29  & Tull \\
2459072.85339  & -48318   &  28  &  Tull \\
2459073.85204   &-48793   &  35  &  Tull \\
2459090.80845   & -49031  & 24   &  Tull \\
2459091.79470   & -48130  &23    &  Tull \\
2459104.75247   &-48831   &  42  & Tull \\
2459114.70318  & -48407 & 31   & Tull \\
2459115.79221  & -48704   & 31 & Tull \\
2459116.89061   & -48600  & 30 &  Tull \\
2459133.80040  & -49039  & 36  &  Tull \\
2459134.83585  &  -48109 & 45  &  Tull \\
2459144.76502 & -48426  & 24   & Tull \\
2459145.76340 &  -48731 & 29   & Tull \\
2459171.74582 & -48147  &  11  &  Tull \\
2459191.61890 & -49023  &  19  & Tull \\
2459192.65273 & -48237   & 33  & Tull \\ 
2459203.59849  &  -49018  & 28 & Tull \\\hline

\end{longtable}



\begin{table*}
	\centering
	\caption{Radial velocities of TOI-2046}
	\label{tab2046}
	\begin{tabular}{lccr} 
		\hline
		\textbf{Date (BJD)} & \textbf{RV (m/s)} & \textbf{$\sigma_{RV} (m/s)$} & \textbf{Instrument}\\
		\hline

       2459099.93401 &   -237 &  10 &  Tull \\
       2459099.94896 &	 -179 &  22 &  Tull \\
       2459099.96417 &	 -110 &  23 &  Tull \\
       2459100.90511 &	 -115 &  26 &  Tull \\
       2459100.92022 &	 -125 &  21 &  Tull \\
       2459100.93531 &	 -131 &  40 &  Tull \\
       2459114.85045 &	 -280 &  15 &  Tull \\
       2459115.83681 &	 -145 &  57 &  Tull \\
       2459116.94014 &	  355 &  41 &  Tull \\
       2459133.87820 &	 -186 &  37 &  Tull \\
       2459134.81185 &	  377 &  17 &  Tull \\
       2459143.73240 &	  307 &  16 &  Tull \\
       2459144.64706 &	 -405 &  33 &  Tull \\
       2459145.69545 &	   55 &  24 &  Tull \\
       2459170.70165 &	  252 &  25 &  Tull \\
       2459171.60259 &	 -442 &  31 &  Tull \\
       2459171.90490 &	  -75 &  21 &  Tull \\
       2459191.64141 &	  346 &  35 &  Tull \\
       2459192.71142 &	 -264 &  30 &  Tull \\
       2459202.61011 &	    7 &  30 &   Tull \\
       2459203.57473 &	  314 &  18 &  Tull \\
       2459203.82064 &	  313 &  15 &  Tull \\
       2459227.69379 &	  346 &  21 &  Tull \\
       2459240.67291 &	   23 &  60 &  Tull \\
       2459120.40138  &    -656  &   171 &   OES \\
       2459160.43763  &     528  &   119 &   OES \\
       2459175.51402  &     258  &   213 &   OES \\
       2459184.48565  &     841  &    94 &   OES \\
       2459188.46491  &     547  &   129 &   OES \\
       2459242.28485  &    1418  &   102 &   OES \\
       2459246.27680  &     607  &   186 &   OES \\
       2459266.25817  &       0  &    85 &   OES \\
       2459269.31186  &     887  &    92 &   OES \\
       2459270.27897  &     326  &   132 &   OES \\
	2459246.26373 & -9419 &	335 &        TCES \\
2459266.45564 & -9047 &	142 &	TCES \\
2459270.52201 & -9147 &	162 &	TCES \\
2459271.30546 & -8563 &	206 &	TCES \\
2459298.29829 & -9403 &	102 &	TCES \\
2459299.42600  & -9237 &	151 &	TCES \\
2459299.55290  & -8613 &	119 &	TCES \\
2459300.29025	 & -9705 &	167 &	TCES \\
2459303.32457	 & -9905 &	138 &	TCES \\
2459303.34573 & -9958 &	232 &	TCES \\
2459304.32230 & -9414 &	97 &	TCES \\
2459304.34403 & -9478 &	121 &	TCES \\
2459305.31905 & -9470 &	103	  &      TCES \\
2459305.34021 & -9538 &	153 &	TCES \\
2459309.32588 &  -10303 &	136 &	TCES \\
2459309.34761 & -10030 &	163 &	TCES \\
2459310.39691 &  -9277 &	291 &	TCES \\
		\hline
	\end{tabular}
\end{table*}


\bsp	
\label{lastpage}
\end{document}